\documentclass[manuscript,screen]{acmart}

\AtBeginDocument{%
  \providecommand\BibTeX{{%
    \normalfont B\kern-0.5em{\scshape i\kern-0.25em b}\kern-0.8em\TeX}}}

\setcopyright{acmlicensed}
\copyrightyear{2018}
\acmYear{2018}
\acmDOI{XXXXXXX.XXXXXXX}

\acmConference[Conference acronym 'XX]{Make sure to enter the correct
  conference title from your rights confirmation emai}{June 03--05,
  2018}{Woodstock, NY}
\acmISBN{978-1-4503-XXXX-X/18/06}

\usepackage{xspace}
\usepackage{amsmath}
\usepackage{graphicx}
\usepackage{booktabs}
\usepackage{colortbl}
\usepackage{multirow}
\usepackage{url}
\usepackage{diagbox}
\usepackage{breakurl}
\usepackage{caption}
\usepackage{subcaption}
\usepackage{bm}
\usepackage{adjustbox}
\usepackage{soul}
\usepackage[ruled, vlined, linesnumbered]{algorithm2e}

\begin{document}
\renewcommand{\arraystretch}{1.2}
\newcommand{\model}{TUEF\xspace}
\title[Leveraging Topic Specificity and Social Relationships for EF in CQA]{Leveraging Topic Specificity and Social Relationships for Expert Finding in Community Question Answering Platforms}

\author{Maddalena Amendola}
\email{maddalena.amendola@phd.unipi.it}
\orcid{0000−0001−6556−4032}
\affiliation{%
  \institution{IIT-CNR, ISTI-CNR}
  \city{Pisa}
  \country{Italy}
}

\author{Andrea Passarella}
\authornotemark[1]
\email{andrea.passarella@iit.cnr.it}
\orcid{0000−0002−1694−612X}
\affiliation{%
  \institution{IIT-CNR}
  \city{Pisa}
  \country{Italy}
}

\author{Raffaele Perego}
\authornotemark[1]
\email{raffaele.perego@isti.cnr.it}
\orcid{0000−0001−7189−4724]}
\affiliation{%
  \institution{ISTI-CNR}
  \city{Pisa}
  \country{Italy}
}

\renewcommand{\shortauthors}{Amendola et al.}

\begin{abstract}
Online Community Question Answering (CQA) platforms have become indispensable tools for users seeking expert solutions to their technical queries. The effectiveness of these platforms relies on their ability to identify and direct questions to the most knowledgeable users within the community, a process known as Expert Finding (EF). EF accuracy is crucial for increasing user engagement and the reliability of provided answers. Despite recent advancements in EF methodologies, blending the diverse information sources available on CQA platforms for effective expert identification remains challenging. In this paper, we present \model, a \textit{Topic-oriented User-Interaction model for Expert Finding}, which aims to fully and transparently leverage the heterogeneous information available within online question-answering communities. \model integrates content and social data by constructing a multi-layer graph that maps out user relationships based on their answering patterns on specific topics. By combining these sources of information, \model identifies the most relevant and knowledgeable users for any given question and ranks them using learning-to-rank techniques. Our findings indicate that \model's topic-oriented model significantly enhances performance, particularly in large communities discussing well-defined topics. Additionally, we show that the interpretable learning-to-rank algorithm integrated into \model offers transparency and explainability with minimal performance trade-offs.  
The exhaustive experiments conducted on six different CQA communities of Stack Exchange show that \model outperforms all competitors with a minimum performance boost of 42.42\% in P@1, 32.73\% in NDCG@3, 21.76\% in R@5, and 29.81\% in MRR, excelling in both the evaluation approaches present in the previous literature.
\end{abstract}



\received{20 February 2007}
\received[revised]{12 March 2009}
\received[accepted]{5 June 2009}

\maketitle

\section{Introduction}
Online Community Question-Answering (CQA) platforms, such as StackOverflow and AskUbuntu, have become indispensable tools for users seeking expert solutions to their technical queries. These platforms are built upon the collaborative efforts of users who pose questions and provide answers, thus creating an extensive repository of shared knowledge. The success of CQA platforms relies on their ability to effectively identify and direct questions to the most knowledgeable experts within the community, thus engaging the users. This crucial process, known as Expert Finding (EF), is vital for ensuring the accuracy and reliability of the answers provided.

The EF task focuses on identifying and recognizing users with a high level of expertise who can offer accurate and timely responses to posted questions. It significantly enhances user engagement, trust,  and satisfaction by routing questions to the most knowledgeable community members. Despite advancements in EF methodologies, there are still challenges in effectively integrating the diverse information sources available on CQA platforms. Current proposals for addressing the EF task in CQA contexts rely on information derived from the textual content of questions and answers and from features and signals that model the users' engagement with this content and their interactions within the networked community. The complexity of such interactions and the dynamic nature of expertise necessitates a comprehensive approach that can adapt to varying contexts and user behaviors, which, to the best of our knowledge, still needs to be fully addressed in the literature. 

In this paper, we comprehensively address the EF task by proposing \model, a \textit{Topic-oriented User-Interaction model for Expert Finding}, which aims to fully and transparently utilize the vast amount of heterogeneous information available within online question-answering communities.
\model integrates content and social data by constructing a topic-based multi-layer graph that maps out user relationships based on their topical answering patterns. By combining these sources of information, \model aims to identify the most relevant and knowledgeable users for any given question.
\model operates by first generating the multi-layer graph, where each layer represents a major topic within the community. These topics are automatically identified through the analysis of tags from past questions. Within each layer, nodes represent active users in topic discussions, and edges model the similarities and relationships among these users. When a question is posed, \model uses the multi-layer graph and a ranking model to determine the relevant topics and corresponding graph layers.
In each relevant layer, \model selects candidate experts from two perspectives: 
(i) \textit{Network Perspective}: identifying central users who have significant influence within the community;
(ii) \textit{Content Perspective}: identifying users who have previously answered similar questions.
Navigating the graph from seed nodes identified according to these criteria, \model explores and collects candidate experts through appropriate exploration policies. It then extracts static and query-dependent features based on text and graph relationships for these candidates. Finally, \model applies Learning-to-Rank techniques to score and rank the candidates by their expected relevance to the question.

This work builds upon a previous paper \cite{tuef}, extending it with the following main new contributions:
\begin{itemize}
\item We introduce a new categorization of state-of-the-art EF models based on the evaluation approach they adopt: \emph{Expert Ranking} and \emph{Expert Subsample Ranking}. Such categorization contributes to the definition of good practices for a fair EF model evaluation.
\item We conduct an ablation study of \model across six scientific communities: StackOverflow, Unix, AskUbuntu, Server Fault, Physics, and Mathematics. This analysis allows us to understand the contribution of the various CQA information sources and \model components to the end-to-end predictive performance and to validate the generality of \model performance.
\item We integrate an interpretable Learning-to-Rank algorithm \cite{lucchese2022ilmart} into the \model framework and provide examples of interpretation, showing how we can make the model fully transparent without losing its accuracy.
\item We compare \model with state-of-the-art competitor models within the \emph{Expert Ranking} category, to which \model belongs. Moreover, we adapt \model for a fair comparison with state-of-the-art models in the \emph{Expert Subsample Ranking} category, accommodating this different evaluation configuration. We thus highlight the overall advantage of \model concerning relevant alternatives in the state of the art. 
\item We study the ability of {\model} to scale with larger datasets by considering four datasets corresponding to 1, 2, 3, and 4 months of StackOverflow data.
\end{itemize}

Our findings indicate that the multi-layer graph significantly enhances performance when the topics discussed in the community involve a well-defined clustering of question tags, as in StackOverflow. In the cases of small communities characterized by a less broad distribution of discussion topics, \model using single- or multi-layer graphs exhibit comparable performance. As discussed in \cite{tuef}, content information consistently emerges as the most crucial component for EF in all communities. Additionally, we show that integrating an interpretable Learning-to-Rank solution into \model typically results in a slight performance decrease. However, particularly in some communities, the reduction in prediction performance is minimal, making the interpretable version of \model a valuable option for gaining insights into the decision-making process. Finally, the results of extensive experiments conducted on CQA communities with varying characteristics show that \model surpasses all baseline models, excelling in both the \emph{Expert Ranking} and \emph{Expert Subsample Ranking} evaluation categories.

The article is structured as follows: in Section \ref{sec:related}, we present state-of-the-art models divided based on two different points of view: (i) the information used (Text-, Feature-, and Network-based methods) and (ii) the evaluation approach adopted (Expert Ranking and Expert Subsample Ranking). Next, Section \ref{sec:method} details the \model framework. Following, in Section \ref{sec:experiments}, we present the experimental setup and detail the communities examined, the settings of \model hyperparameters, and the metrics used to assess its performance. Finally, in Section \ref{sec:results}, we present the results of the \model ablation study across all communities and the comparison between \model and the state-of-the-art baselines under the two evaluation-based categories.
 
\section{Related Work}
\label{sec:related}

In this section, we categorize state-of-the-art models from two perspectives: the type of information the models use (textual, features, and network) and the evaluation methodology the authors adopt for their assessment (with or without subsampling).

\subsection{Information-based classification}
Research proposals in the field of the EF task for CQA platforms can be grouped into three broad groups based on the information they explore to address the task: text-based, feature-based, and network-based methods.

\paragraph{\textbf{Text-based methods}}
A variety of methods have been proposed to address the EF task by leveraging similarities between current and previously answered questions. Dehghan and Ansell \cite{DBLP:journals/tkdd/DehghanA19} introduced a model that clusters question terms based on semantic similarity and co-occurrence. Similarly, Liang et al. \cite{DBLP:conf/www/Liang19} proposed a semantic unsupervised generative adversarial network to assess similarities between word representations and experts.
Moreover, Dehghan et al. \cite{DBLP:journals/ipm/DehghanBA19} modeled user expertise by utilizing the tree structure of various domains, tags, and the temporal dimension of user response behavior. In a related vein, Zhang et al. \cite{DBLP:conf/wsdm/ZhangCZCXLC20} considered the temporal dimension by proposing models with multi-shift and multi-resolution settings to capture temporal dynamics effectively.
Fu et al. \cite{fu2020recurrent} introduced a novel approach using the Recurrent Memory Reasoning Network (RMRN), which employs different reasoning memory cells equipped with attention mechanisms to focus on various aspects of the question. Building on the concept of attention mechanisms, Peng et al. \cite{peng2022towards, peng2022towards_comprehensive, liu2022expertbert, peng2023contrastive} implemented these mechanisms in multi-view, multi-grained, semi-supervised pre-trained, and pre-trained models with personalized fine-tuning for expert finding.
Moreover, Qian et al. \cite{qian2022multi} proposed a Multi-Hop Interactive Attention-based Classification Network (MIACN) that utilizes attention mechanisms to detect latent interactions among question subjects and bodies.

\paragraph{\textbf{Feature-based methods}}

The second group of methods in EF relies on hand-crafted features to model the expertise of community members. Roy et al. \cite{roy2018finding} formalized a scoring function to capture various aspects of expertise, including the propensity to answer questions related to profile tags, the ability to provide accepted answers, and recent activity levels. Mumtaz et al. \cite{mumtaz2019expert2vec} proposed a framework integrating activity, community, and time-aware features, with the temporal aspect also considered in \cite{fu2019tracking} to track and incorporate the evolution of user roles, as well as in \cite{kundu2019finding}.
Similarly, Tondulkar et al. \cite{tondulkar2018get} included features that favor experts providing high-quality answers to complex questions, utilizing a comprehensive set of features capturing user availability and knowledge, and applying Learning-to-Rank (LtR) methods for expert ranking. The quality and consistency of answers are further addressed by Faisal et al. \cite{faisal2019expert}, who proposed a model based on an adaptation of the bibliometric g-index. LtR techniques are further utilized by Sorkhani et al. \cite{sorkhani2022feature}, introducing 74 content-based and social-based features.
An important aspect in addressing the EF task is considering the intimacy between the asker and answerer, studied by Fu et al. \cite{fu2020user}. As in text-based approaches, Tan et al. \cite{tang2020hierarchical} employed attention mechanisms by proposing a method using Hierarchical Attentional Factorization Machines (HAFM). This approach combines factorization machines with hierarchical attention mechanisms to effectively model user-expert interactions and to highlight the importance of various features in expert recommendation tasks. The hierarchical attention mechanism helps in prioritizing relevant information at both the user and expert levels, enabling more accurate and personalized recommendations. Contrary to many studies focusing primarily on identifying the most senior platform users as experts, Roy et al. \cite{roy2024early} aim to predict promising expert users at an early stage in community question answering sites.

\paragraph{\textbf{Network-based methods}}
\textit{Network-based methods} integrate interactions and relationship information within networks. Kundu et al. \cite{kundu2019formulation} defined a framework that comprises a text-based component for assessing expert knowledge on specific topics, accompanied by a Competition Based Expertise Network (CBEN) \cite{aslay2013competition}. This network uses link analysis techniques, such as AuthorRank \cite{liu2005co} and Weighted HITS \cite{li2002improvement}. The CBEN approach is further utilized in \cite{kundu2020preference}, where connections between two users are established if they have responded to the same question. Additionally, this study incorporates both \textit{intra-profile} and \textit{inter-profile} preferences of community users. In \cite{kundu2021topic}, the same authors propose a topic-sensitive hybrid expertise retrieval system (TSHER) that integrates assessments of knowledge, reputation, and authority. Le et al. \cite{le2018retrieving} measure the similarity between answerer and asker using social network techniques, employing Random Walks with Restart to calculate the proximity between two nodes.

Contrarily, Sun et al. \cite{DBLP:conf/icwsm/SunMR018} propose a language-agnostic perspective of user expertise, constructing a competition graph with user and question nodes where edges represent increasing levels of question difficulty, thus depicting the hierarchical structure of questions. This concept of hierarchy is further explored in \cite{DBLP:conf/aaai/SunBBLS019}. Addressing data sparsity, Sang et al. \cite{sang2019multi} introduced a Multi-modal Multi-view Semantic Embedding (MMSE) framework that learns semantic embeddings from both local and global perspectives, incorporating social structure information to enhance embedding quality. Ghasemi et al. \cite{ghasemi2021user} focus on user embedding, proposing a joint model for text and node similarity.

Moreover, Li et al. \cite{li2019personalized} represent the CQA platform as a Heterogeneous Information Network (HIN), aiming to learn embeddings for nodes representing question contents, raisers, and answerers. They employ an LSTM-equipped Metapath-based Embedding algorithm and a CNN scoring function to rank experts. The use of HIN and metapath-based algorithms is also noted in \cite{qian2022heterogeneous}. Liu et al. \cite{liu2022high} apply advanced semantic analysis and interest drift modeling to evaluate experts based on the relevance and depth of their contributions in specific domains. The concept of interest drift is similarly addressed by Krishna et al. \cite{krishna2023temporal}, who propose a simple graph diffusion-based expert recommendation model that accounts for semantic and temporal information. Temporal dynamics in expertise assessment are further examined by Costa et al. in \cite{costa2023here, costa2023ask}, who introduce temporally-discounted, tag-based models.

\subsection{Evaluation-based classification}
The EF task is commonly cast to an item recommendation task in which the items are community users to be ranked by a measure of likelihood they can effectively answer the new question. The users with the highest scores are the ones to whom the new question should be routed.
When evaluating recommender systems, there are two important methods to consider: online evaluation and offline evaluation. Online, end-to-end evaluation is the primary and most reliable way to evaluate a recommender system \cite{hofmann2016online, liang2018variational}, but it is not applicable during the development of a new model \cite{dallmann2021case}. On the other hand, offline evaluation is the main instrument available for academic research and allows for exploring hyperparameter settings \cite{canamares2020target, shani2011evaluating}. It is commonly used to test a new model's effectiveness before moving to online evaluation.
Moreover, in item recommendation tasks, the catalog of items to retrieve from is usually large, and finding matching items from this large pool is challenging \cite{krichene2020sampled}. 
Considering these factors, recent studies have speeded up the evaluation process by calculating the performance metrics of interest on a target set only. This target set is usually a small sample of the items possibly matching the query that includes all relevant items and a defined number of negative (non-relevant) items.
However, in recent years, researchers have analyzed and critically evaluated these sample metric strategies. Krichene et al. \cite{krichene2020sampled} thoroughly investigated sampled metrics, revealing that they are inconsistent with their exact version. This finding is significant as it means that these sampled metrics do not persist relative statements, such as recommender A is better than B, not even in expectation. Furthermore, they found that the smaller the sampling size, the less difference between metrics. Canamares et al. \cite{canamares2020target} found that comparative evaluation using reduced target sets contradicts, in many cases, the corresponding outcome using large targets. Finally, Dallmann et al. \cite{dallmann2021case} explored two widely used sampling strategies, \emph{sampling by popularity} and \emph{uniform random sampling}, and found that both can produce inconsistent rankings compared with the full ranking of the models and consistently produce different ranking when compared over different sample sizes.

When considering the EF context, one effective strategy to simplify the task and reduce complexity is identifying expert users based on specific criteria, such as their consistent provision of high-quality answers. However, it's crucial to acknowledge that this approach can intensify the cold-start problem, potentially leading to the exclusion of users who could be highly relevant to new questions. Despite this, the method significantly accelerates metric computation and enhances system accuracy. 
So far, different strategies have been adopted to define a subset of users representing the experts: 10\% of the most active users \cite{kundu2021topic,qian2022heterogeneous, tang2020hierarchical,peng2022towards,peng2022towards_comprehensive,liu2022expertbert,peng2023contrastive,roy2024early,li2019personalized,kundu2019formulation,liu2022efficient}, all users with the number of accepted answers greater than a threshold \cite{kundu2021time, krishna2023temporal, qian2022multi, sun2018qdee, sun2019atp, fu2020recurrent, fu2020user, kundu2020preference, fu2019tracking}, or a two-stage approach that considers the acceptance ratio \cite{nobari2020quality, fallahnejad2022attention, kasela2023se, dargahi2017skill, mumtaz2019expert2vec}.
No specific heuristic can determine whether a user should be considered an expert. However, to conduct a fair comparison among different models for the EF task, it is important to distinguish studies according to the sampling approach adopted during the ranking phase, ensuring that the models are compared under the same assumptions. Specifically, we recognize two different methods:
\begin{itemize}
    \item \textbf{Experts Ranking}: For a new question, this group of works essentially ranks all users labeled as \emph{expert users} in the first stage. The limitation of this approach during the evaluation phase is that it requires the test set to consist of questions for which the best answerer is a user who was previously labeled as an expert. On the other hand, this evaluation methodology provides realistic figures of the end-to-end performance of the observed system. The following works belong to this group: \cite{nobari2020quality,kundu2021topic,fallahnejad2022attention,liu2022high,costa2023here,kundu2021time,costa2023ask,krishna2023temporal,qian2022multi,roy2024early,sun2019atp,fu2019tracking,fu2020recurrent,mumtaz2019expert2vec,kundu2019formulation,sun2018qdee,fu2020user,kundu2020preference}. 
    \item \textbf{Experts Subsample Ranking}: Given the new questions along with the information of all the users who actually answered the question (ground truth), the studies in this group rank only a fixed-size subsample of users. 
    This set of users is generally not large and always includes the ground truth users plus some other users who didn't answer the question (negative examples).  In most of the works following this approach the size of the subsample is 20 and negative examples chosen randomly among the 10\% most active users are added to the subsample until it reaches the desired cardinality. While this approach overcomes  one limitation of the \textit{Experts Ranking} methodology by requiring only the knowledge of the answers of the questions in the training set, it relies on a sampling strategy that might produce inconsistent rankings if the sample size is varied and can result in unreliable system rankings\cite{krichene2020sampled,canamares2020target,dallmann2021case}. The works belonging to this group include the following: \cite{sang2019multi,ghasemi2021user,qian2022heterogeneous,tang2020hierarchical,sorkhani2022feature,peng2022towards,peng2022towards_comprehensive,liu2022expertbert,peng2023contrastive,li2019personalized,DBLP:conf/wsdm/ZhangCZCXLC20,liu2022efficient}.
\end{itemize}

\begin{figure}
    \centering
    \includegraphics[width=0.9\linewidth]{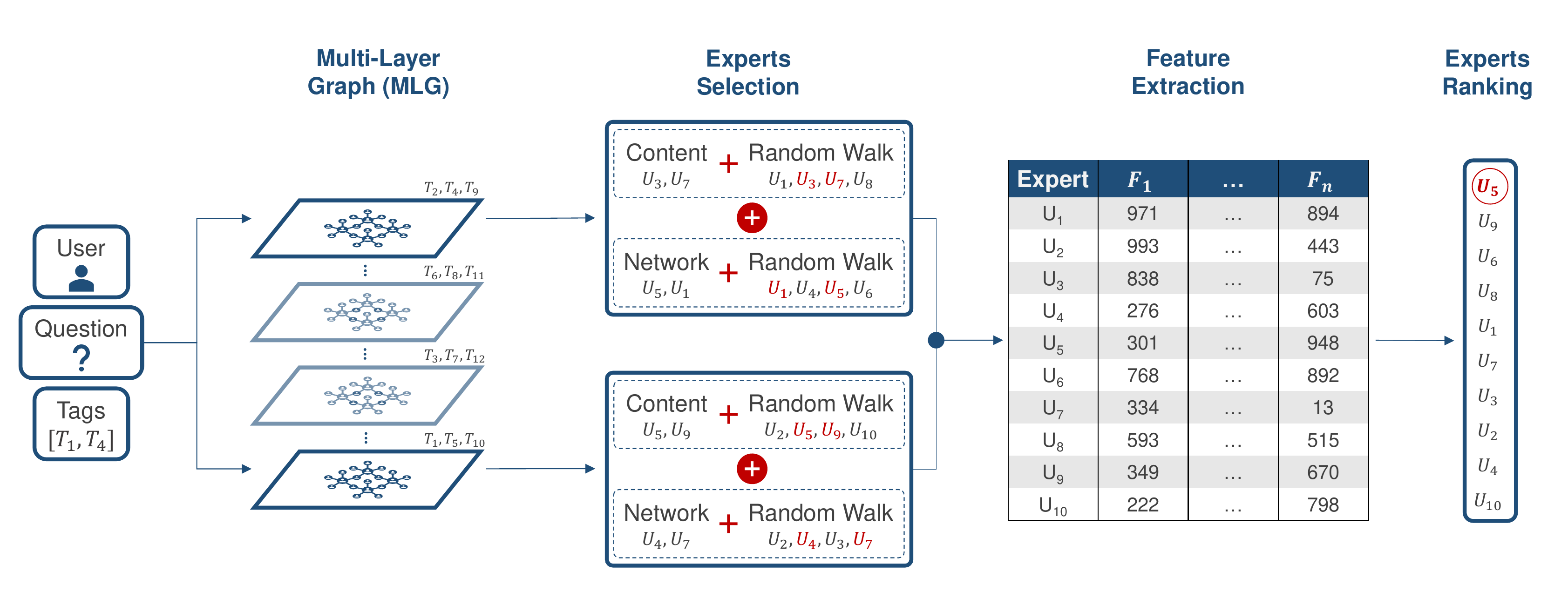}
    \caption{Illustration of the \model approach highlighting the distinct components. At inference time, \model first determines the main topics to which the question $q$ belongs and the corresponding graph layers (Multi-Layer Graph). Next, for each layer, it selects the candidate experts from two perspectives: i) \textit{Network}, by identifying central users that may have considerable influence within the community; ii) \textit{Content}, by identifying users who previously answered questions similar to $q$. The Multi-Layer Graph is used to collect candidate experts through Random Walks (Expert Selection). Following, \model extracts features based on text, tags, and graph relationships for each selected experts (Feature Extraction). Finally, \model uses a learned, precision-oriented model to score the candidates and rank them by expected relevance (Experts Ranking).}
    \label{fig:architecture}
\end{figure}

\section{Topic-oriented User-interaction model for Expert Finding (TUEF)}
\label{sec:method}

The \model framework leverages topic-oriented content similarity and user relationships to facilitate the selection and ranking of experts. This approach is based on two fundamental considerations:
\begin{itemize}
\item CQA platform users utilize tags to characterize their questions and increase the likelihood of receiving relevant answers. The questions' tags help identify topical areas within the community relevant to specific questions.
\item Unlike traditional social networks, user relationships on CQA platforms are not explicitly defined. However, implicit relationships can be discerned by analyzing users' question-answering behaviors. The knowledge gained from these interactions can enhance answer quality and user engagement with the platform.
\end{itemize}

Figure \ref{fig:architecture} illustrates the logical organization of \model.
To effectively integrate the valuable information derived from tags, content, and user relationships, \model adopts a multi-layer graph representation (Section \ref{sec:mlg}) that encapsulates the macro topics discussed within the online community. Each layer of this graph models user relationships at the level of specific macro topics, considering their similarities based on tags, questions, and answering behaviors. \model identifies community \emph{experts}—users known for consistently providing accepted answers—and implements an exploratory algorithm that fully exploits the multi-layer graph structure to select a group of candidate experts for each relevant macro topic associated with the current question (Section \ref{sec:selection}).

This exploratory algorithm integrates social and content-based information, considering experts' centrality within the network and their expertise in the topics of interest. Central users often represent individuals who have garnered significant attention and engagement within the community. Similarly, users close to the query are more likely to connect with experts relevant to the specific topic of interest. By strategically exploring the multi-layer graph starting from these nodes, the algorithm increases the probability of selecting appropriate experts early in the process.

Finally, \model extracts features that represent the identified candidate experts and applies Learning-to-Rank techniques \cite{10.1561/1500000016} (Section \ref{sec:ranking}) to rank them based on their expertise and likelihood of effectively answering the question. This systematic approach ensures the selection of high-quality experts tailored to the specific information needs of the user asking the question.

\subsection{User Interaction Model}
\label{sec:mlg}

In \model, user relationships related to each specific topic are independently modeled. From here on, let $U$ denote the set of active users on the CQA platform under consideration, and let $Q$ represent the set of past questions posted and answered by users in $U$.

\paragraph{\textbf{Topic Identification.}}\label{sec:clustering}
\model employs a tag clustering approach to identify the main topics discussed on the CQA platform based on past questions. The clustering approach, discussed in \cite{clustering}, utilizes the \textit{k}-means algorithm \cite{macqueen1967some} on a tag co-occurrence matrix $M$.
Tags' co-occurrence patterns are leveraged to group semantically related tags, as tags often associated with the same questions tend to be conceptually similar \cite{4597030}. For each question $q \in Q$, the tags associated with $q$ are denoted as $tags(q)$. Let $T$ be the set of all unique tags across questions in $Q$, and let $F$ represent the set of the most frequent tags (top $\lambda$) that serve as clustering features.
The co-occurrence matrix $M^{|T| \times |F|}$ is constructed as follows:
\begin{equation}
    m_{i, j} = |\{q \in Q \;|\; \{t_i, f_j\} \subseteq tags(q), \; t_i \in T, \; f_j \in F\}|
\end{equation}
Here, $m_{i, j}$ indicates the number of questions in $Q$ where the $i_{th}$ tag and the $j_{th}$ feature co-occur. Finally, the matrix rows are normalized to express the relative frequency of tag co-occurrences with each feature. For a given tag $t_i$ and feature $f_j$, the normalized matrix $\hat{M}$ is defined as:
\begin{equation}
    \hat{m}_{i,j} = \frac{m_{i, j}}{\sum_{n=1}^{|F|}{m_{i, n}}}
\end{equation}

Clustering the tag co-occurrence matrix to devise macro topics is based on the observation that CQA users often pair broad, common tags with more specific ones to achieve greater precision in topic identification. By considering the co-occurrences of specific tags with broader tags, we can enhance the categorization of questions. Using the normalized matrix $\hat{M}$ and a specified value of \textit{k} for the \textit{k}-means algorithm, the clustering process yields \textit{k} disjoint clusters of tags representing primary community domains. The optimal \textit{k} value is determined based on silhouette maximization criteria \cite{rousseeuw1987silhouettes}, as described in Section~\ref{sec:experiments}. This comprehensive approach to topic identification and clustering ensures that \model effectively captures and represents the main topical areas discussed within the online community, facilitating a good understanding of community dynamics and interactions.

\paragraph{\textbf{Multi-Layer Graph.}}
\model meticulously models user relationships within each layer by treating users $U$ as nodes in a multi-layer graph and establishing connections based on similar patterns of providing accepted answers within specific layers. 
Formally, the structure of \model is defined using a multi-layer graph $G = [L_1, ..., L_k]$, where each layer $L_i$ corresponds to a tag cluster. Each layer $L_i = (V_i, E_i)$ is an independent graph, with nodes $V_i$ representing users associated with the layer and edges $E_i$ denoting their relationships. 
To ensure consistent and accurate answers within a specific layer, \model includes in $V_i$ only those users who have provided a number of accepted answers equal to or exceeding a given $\epsilon$ percentile. 
Moreover, to characterize users' knowledge in $V_i$, each user $u \in V_i$ is assigned a topic vector $b^i_u$. This vector represents the user's contribution to various tags within the layer, normalized by their total contribution across all layers.
Specifically, for a tag $t_j$ within $L_i$, the $j_{th}$ position of $b^i_u$ is calculated as:
\begin{equation}
    b^i_u[j] = \frac{accepted(L_i, u, t_j)}{\sum_{\forall z,m}  accepted(L_z, u, t_m)}
\end{equation}

Here, $accepted(L_i, u, t_j)$ denotes the number of accepted answers provided by user $u$ for questions labeled with tag $t_j$ in layer $L_i$. Finally, after computing the topic vectors for all the users in $V_i$,  the cosine similarity is calculated between each pair of users within the layer. If the similarity between two users $u_a$ and $u_b$ exceeds a predefined threshold $\delta$, an edge $(u_a, u_b)$ weighted by the cosine similarity value is added to $E_i$.

It is crucial to emphasize that each question in \model is associated with tags assigned to specific layers within $G$. Consequently, a question can belong to multiple layers, and users answering these questions are represented across all associated layers. This multidimensional representation of user expertise and relationships across layers enables \model to capture a comprehensive view of users' knowledge and interactions within the CQA platform. This approach ensures that users' expertise and social connections transcend individual layers, reflecting the nature of community participation and expertise within the platform.

\subsection{Expert Selection}
\label{sec:selection}

The \emph{Expert Selection} component, as depicted in Algorithm \ref{alg:find_experts}, has the goal of selecting for each new question $q$ submitted to the CQA platform a set of users who are likely to be highly knowledgeable of the topics of $q$, i.e., the set of candidate experts. The component operates iteratively for each layer to which question $q$ belongs. This selection process is structured into three distinct phases: (i) \textit{Sorting phase:}, where the nodes within each layer are arranged in a specified order to facilitate subsequent expert selection (line 2); (ii) \textit{Collection phase:}, which selects an initial set of candidate experts from the sorted node lists (line 4); (iii) \textit{Exploratory phase:}, that expands the initial set by exploring the graph structure to identify additional potential experts (line 6).
The remainder of this section investigates the Expert Identification process, which classifies users $u \in U$ as either \emph{experts} or \emph{non-experts}, followed by a detailed exploration of the three phases mentioned above.

\begin{algorithm}
\caption{Expert Selection}\label{alg:find_experts}
\SetKwInOut{Input}{Input}\SetKwInOut{Output}{Output}
\Input{the layer $L_i$, the question $q$, \\ the method $m$ (Network or Content),\\
the probability $p$}
\Output{the set $D_i$ of candidate experts for $q$ in $L_i$ }
//sort $L_i$ nodes based on the method criteria\;
$sorted\_nodes \leftarrow sort\_nodes(L_i, \:q, \:m)$\;
//select the initial set of experts based on probability\;
$initial\_set \leftarrow reach\_probability(sorted\_nodes, p)$\;
//expand the initial set by exploring the graph\;
$D_i \leftarrow explore\_graph(L_i, \:initial\_set)$\;
\Return{$D_i$}
\end{algorithm}

\paragraph{\textbf{Expert Identification}}
\label{sec:exp_rep}

Experts in CQA platforms can be identified heuristically by considering the number of \textit{accepted answers} they provided in the past. 
Another signal of user trust is given by the acceptance ratio $r_u$, i.e.,  the ratio between the number of accepted answers and the total number of answers a given user provides. 
As in \cite{dargahi2017skill}, in \model, we adopt an expert selection criterion based precisely on these two measures: first, we select the set $C \subseteq U$ of \textit{candidate experts} by considering all the users having a number of accepted answers greater or equal to a specified threshold $\beta$: next, we label as \textit{experts} the set $E \subseteq C$ of users whose acceptance ratio $r_u$ is greater than the overall average $\overline{r}$:
\begin{equation}
    E = \{ u \in C \; | \; r_u > \overline{r} \},\quad \overline{r} = \frac{\sum_{u \in C}{r_u}}{|C|}
\end{equation}
Each layer within the multi-layer graph comprises nodes labeled as \textit{experts} and \textit{non-experts}. While the former are the primary targets of the Expert Selection process, the latter facilitates comprehensive exploration of the graph structure.

By implementing this process, we aim to identify users who consistently provide valuable and accepted solutions within the community, as they are likely to be proficient candidates for effectively addressing new questions posted on the platform.

\paragraph{\textbf{Sorting}}
\model adopts a comprehensive approach to identify candidate experts for a given question $q$ by leveraging content and social information. This process involves two key perspectives: the \textit{Network-based} perspective, focusing on users' centrality within the network, and the \textit{Content-based} perspective, assessing users' relevance to the newly posted query based on their past interactions.
In the Network-based approach, \model utilizes Betweenness centrality \cite{freeman1977set}, which quantifies the centrality of nodes within a graph by evaluating their influence over information flow. Nodes with higher Betweenness centrality are Considered more central within the network. Each node $v_j^i$ in layer $L_i$ is assigned a Betweenness score $s_j^i$, which is used to sort nodes in descending order.
On the other hand, the Content-based approach involves sorting layer nodes according to their similarity to the query $q$ based on questions previously answered by experts. This is achieved using Information Retrieval techniques and pre-built indexes, specifically the \textit{TextIndex} and \textit{TagIndex}, which index the text and associated tags of historical questions, respectively. When a new question is presented with its tags, the Content-based method employs a retrieval model (we specifically use BM25) independently on both indexes to retrieve a sorted list of relevant questions, along with information about the experts who provided accepted answers. The query for the \textit{TextIndex} includes the concatenated question, title, and body, while for the \textit{TagIndex} it consists of the concatenated question tags. Subsequently, the lists retrieved from each index are merged alternately, preserving the original order of elements.

This approach results in a query-specific ordering of nodes within each layer $L_i$, effectively leveraging network centrality and content relevance to identify and rank candidate experts tailored to the context of the new query $q$. The combination of these perspectives enhances the precision and relevance of expert selection within the \model framework.

\paragraph{\textbf{Candidate Collection}}

In the candidate collection phase, the objective is to select a subset of experts $D_i \subseteq V_i$ from each layer $L_i$. This process balances the need for a sufficiently large candidate pool to ensure high recall (the probability of obtaining the correct answer) with the desire to include relevant experts for high precision.
To achieve a balance between precision and recall, we estimate the probability $p$ of not receiving an answer from any user in $D_i$. Initially, $p$ is set to 1 when $D_i$ is empty and is incrementally reduced as experts are added to $D_i$. Each expert $u$ added to $D_i$ contributes to lowering $p$ based on their acceptance ratio $\mu_u$, defined as the ratio of accepted answers to total answers provided by the user. Furthermore, to refine the modeling of topic-based expertise, we smooth $\mu_u$ by considering the user's activity within the specific layer: the smoothing factor is computed by adjusting $\mu_u$ based on the ratio of expert answers within the layer to the maximum number of answers provided by any user within that layer.
The probability $p$ is updated iteratively using the formula:
\begin{equation}
    p = p \cdot (1 - \mu_u)
\end{equation}
This iterative process continues until $p$ becomes less than or equal to a predefined threshold $\alpha$. Once the threshold $\alpha$ is met, the inclusion of new candidates to $D_i$ is halted, and the exploratory phase of the expert selection process begins. This strategic approach ensures that the candidate pool $D_i$ comprises experts likely to provide valuable and accurate answers to the specific question while representing good starting points for MLG exploration.

\paragraph{\textbf{Exploratory Phase}}

In the Exploratory phase, \model explores the graph structure to identify additional candidate experts that may not have been identified with the previously described phases, specifically exploiting the implicit relationships between users, aiming to enhance recall. 
The starting point for this exploration is the set of experts $D_i$ from each layer $L_i$.
Specifically, for each node $v_i \in D_i$, \model initiates a series of fixed-length Random Walks. At each step of the Random Walk, the next node to visit is selected randomly based on a probability distribution $d$, computed considering the neighbors of the current node $v_i$ and their link weights, representing the similarities with neighboring nodes. 

More formally, let $\mathcal{N}i = {v{i1}, \ldots, v_{iN_i}}$ denote the neighbors of expert node $v_i$, where $N_i = |\mathcal{N}_i|$ is the total number of neighbors of $v_i$.
The probability $d_j$ of visiting a neighbor node $v_{ij}$ is calculated as:
\begin{equation}
    d_j = \frac{w_{ij}}{\sum_{z=1}^{|N_i|}{w_{iz}}}
\end{equation}
Here, $w_{ij}$ represents the weight of the edge between node $v_i$ and its neighbor $v_{ij}$. 
Nodes with stronger connections to the current node $v_i$ are more likely to be selected during the Random Walks. Whenever an expert node is encountered during the exploration, it is added to the current set of candidate experts if not already included.

This exploratory process is carried out independently for each layer to which the question belongs, ensuring a comprehensive exploration of both Network-based and Content-based perspectives within each layer. Combining Random Walks with tailored probability distributions enriches the candidate pool $D_i$ with potentially neglected experts, enhancing the overall recall and effectiveness of the expert selection process in the \model framework.

\subsection{Ranking candidate experts}
\label{sec:ranking}
\model leverages \textit{Learning to Rank} (LtR) algorithms \cite{10.1561/1500000016} to learn an effective ranking function from training data. LtR methods exploit a labeled dataset to learn a scoring function $\sigma$ that approximates the ideal ranking function inherent in the training examples.
\model selects a subset of past questions, used to model user relationships as discussed in Section \ref{sec:mlg}, to serve as the training set $T \subset Q$ for the LtR algorithm. Each query $q$ in the training set $T$ is associated with a set of candidate experts $CE \subset U$. Moreover, for each query-candidate pair $(q, u_i)$, where $q \in T$ and $u_i \in CE$, there exists a \emph{relevance judgment} $l_i$ that indicates whether $u_i$ is an expert for query $q$.
Each query-candidate pair $(q, u_i)$ is represented by a feature vector $x$ that encapsulates information about the query, the candidate expert, and their relationship. The LtR algorithm learns a function $\sigma(x)$ that predicts a relevance score for the input feature vector $x$. This learned function $\sigma(x)$ is subsequently used during inference to compute scores for candidate experts and rank them accordingly.

The features used to model the query and candidate expert can be categorized into two groups: \textit{Static} features and \textit{Query-dependent} features. The group of Static features comprises those that remain the same for each query: the \textit{Reputation} of the expert, the number of \textit{Answers} and \textit{AcceptedAnswers}, the \textit{Ratio} (i.e., the ratio of answers to accepted answers), and \textit{AvgActivity} and \textit{StdActivity}, representing the average and standard deviation of time intervals between consecutive answers provided by the expert.
The Query-dependent features, instead, are computed every time for each query and they include:
\begin{itemize}
    \item \textit{LayerCount}: Number of distinct graph layers in which the expert is selected during the Expert Selection process.
    \item \textit{QueryKnowledge}: Ratio of answers to accepted answers provided by the expert in layers relevant to the query $q$.
    \item \textit{VisitCountContent} and \textit{VisitCountNetwork}: Total number of times an expert is encountered in the Collection and Exploratory phases using Content-based and Network-based techniques.
    \item \textit{StepsContent} and \textit{StepsNetwork}: Number of steps required to discover the expert during Collection or Exploratory phases.
    \item \textit{BetweennessPos} and \textit{BetweennessScore}: Expert's rank in the list of users ordered by Betweenness score and the Betweenness score itself.
    \item \textit{ScoreIndexTag} and \textit{ScoreIndexText}: Sum of BM25 scores of historical questions answered by the expert in IndexTag and IndexText, respectively.
    \item \textit{FrequencyIndexTag} and \textit{FrequencyIndexText}: Number of distinct questions answered by the expert returned by the respective indexes.
    \item \textit{Eigenvector}, \textit{PageRank}, \textit{Closeness}, \textit{Degree}, \textit{AvgWeights}: Centrality measures and network features computed based on the graph layers relevant to the query.
\end{itemize}

Specific rules are applied to aggregate feature values for candidate experts selected in multiple layers. Network-based features (e.g., \textit{BetweennessScore}, \textit{Closeness}, \textit{PageRank}) consider maximum values, while Content-based features (e.g., \textit{FrequencyIndex}, \textit{ScoreIndex}, \textit{QueryKnowledge}) utilize summation. Additionally, certain features like \textit{StepsContent}, \textit{StepsNetwork}, and \textit{BetweennessPos} use minimum values to characterize the candidate experts effectively within the ranking framework. These features collectively contribute to the comprehensive ranking process in \model, accommodating both static and query-specific attributes of candidate experts for optimal ranking performance.

\section{Experimental Setup}
\label{sec:experiments}

The following section presents the experimental setup and details the datasets used, the settings of the \model hyperparameters, and the metrics used to assess performance. 

\subsection{Datasets}
We selected six real-world datasets from the most extensive communities on the StackExchange\footnote{\url{https://stackexchange.com/}} platform, namely StackOverflow, Unix, AskUbuntu, ServerFault, Physics, and Mathematics. These datasets are publicly accessible\footnote{\url{https://archive.org/details/stackexchange}} and encompass all the questions and answers posted by StackExchange users within their respective communities. Each question has a title, a body, and a list of tags. Moreover, the dataset for each community includes all the posted answers for each question. Besides question and answer content, we know the positive or negative votes received by a question or an answer, the number of views, the number of users that selected a given question as a favorite one, and the comments that other users might have written under a question or an answer. Questions can be categorized as \textit{closed}, indicating the presence of an \textit{accepted answer}, or \textit{open}, which may have no answers. Information about the user who posted a question or an answer is available. 
Since an answer can be labeled as accepted only by the user who posted the originating question, the authors of accepted answers provide the human-assessed ground truth for the EF task. Additionally, each participating user in the community has associated additional information, such as a \textit{Reputation score}, which is also useful for the EF task.

\subsection{Parameters setting}
We pre-processed the data to ensure a good set of questions and answers. Moreover, the different steps of \model require setting some parameters detailed below.

\paragraph{\textbf{Pre-processing.}} We pre-processed the data to ensure a good set of questions and answers. Moreover, the different steps of \model require setting some parameters detailed below. We applied established data cleaning procedures, as outlined in \cite{mumtaz2019expert2vec}, to ensure the quality of our dataset. We opted for an eight-year timespan for most communities, excluding StackOverflow and Mathematics. The decision to use a shorter timeframe for these two communities arises from their high daily question volumes, ensuring a more balanced representation across all communities. This strategy aims to maintain a roughly equivalent dataset size for each community.
We removed questions and answers without a specified question's \textit{ID} and \textit{OwnerUserID} (i.e., the ID of the user who posted the question). Subsequently, we retained questions with an \textit{AcceptedAnswerID} (i.e., closed questions) and answers with a valid \textit{ParentID} (i.e., the respective \textit{question ID}). Questions where the asker and the best answerer were the same user were also excluded. We split the dataset, allocating 80\% for training/validation and the remaining 20\% for the test set. Notably, we maintained the chronological order of the questions to preserve temporal integrity. Comprehensive statistics for the resulting datasets are provided in Table \ref{tab:stats}.

\paragraph{\textbf{User Interaction Model.}} To identify the primary topics discussed in the community, we employed the clustering technique outlined in Section \ref{sec:mlg}, focusing on the tags associated with questions in the training set. Considering the top $\lambda=10$ most frequent tags as features, we determined the optimal number of clusters within the range $K=[2, 10]$ that maximizes the Silhouette score. Each cluster represents a specific macro-area, corresponding to a layer in the MLG, which captures interactions among users who have provided a number of accepted answers equal to or exceeding the $\epsilon=90th$ percentile. Users are represented by their topic vectors used to compute the pair-wise cosine similarities  (see Section \ref{sec:mlg}). When modeling relationships, we retained edges with a similarity equal to or greater than $\delta=0.5$.

\paragraph{\textbf{Expert Selection}.} We identify as experts the users with a number of accepted answers greater than or equal to the $\omega=95th$ percentile of the distribution, following the procedure detailed in Section \ref{sec:selection}. The minimum number of accepted answers and the total number of users labeled as experts for each dataset are outlined in Table \ref{tab:stats} under the columns \textit{MinAccAns} and \textit{Experts}, respectively. TagIndex and TextIndex return the 1,000 most similar past questions for each query in the expert selection process, respectively. For both Network and Content methods, after node sorting, the collection phase identifies the initial set of experts $D$ that reaches a probability threshold of $p=0.001$, representing the probability of not receiving an answer. Subsequently, starting for each selected expert in $D$, we conduct 5 Random Walks of, at most, 10 steps.

\paragraph{\textbf{Ranking.}} We utilize the LightGBM\footnote{\url{https://lightgbm.readthedocs.io/en/stable/}} \cite{ke2017lightgbm} implementation of LambdaMART \cite{burges2010ranknet} to learn the \model ranking model. 
The LtR training set is constructed using queries from the training dataset where the accepted answerer is labeled as an expert, including  $0 < \zeta \leq 50,000$ queries ordered from the most recent to the oldest. It is important to note that this is a TUEF parameter: if the dataset contains more than 50,000 queries answered by experts, TUEF will take the latest 50,000; otherwise, it will consider only the available queries. 
After performing the MLG exploration for each query of the LtR training set, we excluded all the queries for which \model couldn't include in the candidate set the expert who provided the accepted answer. For the remaining queries, we extracted features for each query-candidate expert pair, as outlined in Section \ref{sec:ranking}. 
The training set was split into training and validation sets following an 80/20 split. Hyper-parameter tuning is carried out using MRR on the validation set, leveraging the HyperOpt library \cite{bergstra2013making}, and optimizing four learning parameters: $learning\_rate\in[0.0001, 0.15]$, $num\_leaves\in[50, 200]$, $n\_estimators\in[50, 150]$, $max\_depth\in[8, 15]$, and $min\_data\_in\_leaf\in[150, 500]$. 

Table \ref{tab:stats} shows the number of queries used for the LtR and the average length of expert lists to rank under the columns \textit{LtR} and \textit{AvgList}, respectively.

\begin{table}[!ht]
\caption{Statistics of the StackExchange communities used for the experiments.}
\begin{adjustbox}{width=\textwidth}
\begin{tabular}{lcccccccccc}
 & \multicolumn{1}{c}{\textbf{Time Period}} & \multicolumn{1}{c}{\textbf{Tags}} & \multicolumn{1}{c}{\textbf{Clusters}} & \multicolumn{1}{c}{\textbf{Silhouette}} & \multicolumn{1}{c}{\textbf{Train}} & \multicolumn{1}{c}{\textbf{Test}} & \multicolumn{1}{c}{\textbf{MinAccAns}} & \textbf{Experts} & \textbf{LtR} & \textbf{AvgList} \\ \hline
\multicolumn{1}{l}{\textbf{StackOverflow}} & \begin{tabular}[c]{@{}l@{}}2020-12-01\\ 2021-01-01\end{tabular} & 5365 & 10 & 0.805 & 39581 & 9521 & 11 & 350  & 8008 & 111\\ \cline{2-11} 
\multicolumn{1}{l}{\textbf{Unix}} & \multicolumn{1}{c}{\begin{tabular}[c]{@{}l@{}}2015-01-01\\ 2023-09-04\end{tabular}} & \multicolumn{1}{c}{1876} & \multicolumn{1}{c}{6} & \multicolumn{1}{c}{0.463} & \multicolumn{1}{c}{51995} & \multicolumn{1}{c}{12985} & \multicolumn{1}{c}{21} & 183 & 22098 & 118 \\ \cline{2-11} 
\multicolumn{1}{l}{\textbf{AskUbuntu}} & \multicolumn{1}{c}{\begin{tabular}[c]{@{}l@{}}2015-01-01\\ 2023-09-04\end{tabular}} & \multicolumn{1}{c}{2021} & \multicolumn{1}{c}{9} & \multicolumn{1}{c}{0.391} & \multicolumn{1}{c}{43469} & \multicolumn{1}{c}{10166} & \multicolumn{1}{c}{9} & 265 & 16918 & 129 \\ \cline{2-11} 
\multicolumn{1}{l}{\textbf{Server Fault}} & \multicolumn{1}{c}{\begin{tabular}[c]{@{}l@{}}2015-01-01\\ 2023-09-04\end{tabular}} & \multicolumn{1}{c}{2137} & \multicolumn{1}{c}{10} & \multicolumn{1}{c}{0.419} & \multicolumn{1}{c}{30797} & \multicolumn{1}{c}{7398} & \multicolumn{1}{c}{15} & 180 & 10041 & 102\\ \cline{2-11} 
\multicolumn{1}{l}{\textbf{Physics}} & \multicolumn{1}{c}{\begin{tabular}[c]{@{}l@{}}2015-01-01\\ 2023-09-04\end{tabular}} & \multicolumn{1}{c}{811} & \multicolumn{1}{c}{8} & \multicolumn{1}{c}{0.406} & \multicolumn{1}{c}{53412} & \multicolumn{1}{c}{13860} & \multicolumn{1}{c}{31} & 146 & 19192 & 102\\ \cline{2-11} 
\multicolumn{1}{l}{\textbf{Mathematics}} & \multicolumn{1}{c}{\begin{tabular}[c]{@{}l@{}}2022-01-01\\ 2023-09-04\end{tabular}} & \multicolumn{1}{c}{1351} & \multicolumn{1}{c}{9} & \multicolumn{1}{c}{0.499} & \multicolumn{1}{c}{46190} & \multicolumn{1}{c}{11649} & \multicolumn{1}{c}{41} & 141 & 15128 & 85 \\ \hline
\end{tabular}
\end{adjustbox}
\label{tab:stats}
\end{table}

\subsection{Evaluation metrics} 
We use Precision@1 (P@1), Normalized Discounted Cumulative Gain @3 (NDCG@3), Mean Reciprocal Rank (MRR), and Recall@5 (R@5) as our evaluation metrics. The cutoffs considered are short, as finding the relevant results at the top of the ranked lists for the EF task is essential. P@1 assumes a pivotal role in the EF task, emphasizing the primary objective of identifying the most suitable answerer for a given query. NDCG@3 and MRR metrics serve as valuable tools when models exhibit identical P@1 scores, offering a detailed comparison by considering the actual position of the best answerer in the list. R@5, in turn, measures the model's efficacy in locating the best answerer within the top five positions.
Performance metrics and statistical significance tests are computed using the RanX Library \cite{bassani2022ranx}.

\paragraph{\textbf{Reproducibility}} \model is prototyped in Python 3.8.17. All experiments are conducted on an Intel(R) Xeon(R) Platinum 8164 CPU 2.00GHz processor with 503GB RAM on Linux 5.4.0-153-generic. The source code of \model is publicly available\footnote{\url{https://github.com/maddalena-amendola/TUEF}}.

\section{Experimental analysis}
\label{sec:results}

The following Section discusses two blocks of experiments. The first block (Section \ref{sec:ablation}) regards an ablation study of \model across all the communities to understand the contribution to the overall performance of \model's different components (detailed in Section \ref{sec:method}). To make EF decisions transparent and interpretable, the experiments also include assessing the impact of the integration in \model of the interpretable LtR algorithm (IlMart \cite{lucchese2022ilmart}) along with examples of \model's decision-making process.
In the second block of experiments (Section \ref{sec:baselines}), we conduct a comprehensive evaluation of \model aimed at assessing its effectiveness in two scenarios:  end-to-end \textit{Expert Ranking} scenario, and an offline \textit{Expert Subsampling Ranking} scenario, where we conduct experiments with sampled metrics \cite{KDD20}, ranking only a small set of candidates for each query.
In both categories, we compare \model with state-of-the-art competitors for which the implementation is publicly available.
Finally, in Section ~{\ref{sec:scalability}}, we study the ability of {\model} to scale with larger datasets by considering four datasets corresponding to 1, 2, 3, and 4 months of StackOverflow data.

\subsection{Ablation study}
\label{sec:ablation}
We compare \model in an end-to-end scenario with variants of the proposed solution, each exploiting only a subset of the \model components. By examining these different configurations, we can assess and quantify the impact of each component and gain insights into the effectiveness of combining social and content information in our approach for addressing the EF task for CQA platforms: 
\begin{itemize}
    \item \textbf{BC}: It uses the MLG and sorts the experts in the layers related to the new question based on their Betweenness centrality score.
    \item \textbf{BM25}: It uses the MLG and sorts the experts in the question's layers based on the BM25 score computed between the query and their previously answered questions. As in \model, the ranked lists of candidate experts retrieved for the tag and content indexes are merged by alternating their elements.
    \item \textbf{\model\textsubscript{NB}}: It only uses the MLG and applies the Network-based method. Candidates are ranked using a LtR model exploiting the static and the following query-dependent features: LayerCount, QueryKnowledge, VisitCountNetwork, StepsNetwork, BetweennesPos, BetweennessScore, Eigenvector, PageRank, Closeness, Degree, and AvgWeights.
    \item \textbf{\model\textsubscript{CB}}: It only uses the MLG and applies the Content-based method. Candidates are ranked using a LtR model exploiting the static and the following query-dependent features: LayerCount, QueryKnowledge, VisitCountContent, StepsContent, ScoreIndexTag, ScoreIndexText, FrequencyIndexTag, FrequencyIndexTex, Degree, and AvgWeights.
    \item \textbf{\model\textsubscript{SL}}: In contrast to \model, it represents users' relationships in a graph with a Single Layer. All other phases are unchanged.
    \item \textbf{\model\textsubscript{NoRW}}: In contrast to \model, it skips the Exploratory phase and does not perform Random Walks to extend the set of candidate experts selected considering the probability of receiving an answer. 
    \item \textbf{\model\textsubscript{Lin}}: It ranks the experts according to the solution of \cite{roy2018finding}, which uses a linear combination of features modeling experts, questions, and users' expertise.
    \item \textbf{\model\textsubscript{IlMart}}: It uses IlMart, the interpretable version of LambdaMart algorithm proposed in \cite{lucchese2022ilmart} to learn the expert ranking model. IlMart is a learning algorithm that generates explainable LTR models based on a version of LambdaMART constrained to use only univariate and bivariate functions at training time. The IlMart learning strategy is made of three steps: i) \textit{Main Effects Learning}, which learns with the LambdaMART boosting strategy an ensemble of trees $\tau(x_j)$ in which the only feature allowed for each split is $x_j$; ii) \textit{Interaction Effects Selection}, which selects the top-K most important feature pairs $(x_i,x_j)$; iii) \textit{Interaction Effects Learning}, which learns with LambdaMART an ensemble of trees $\tau_{ij}(x_i,x_j)$, where the only features allowed in a split of each tree $t \in \tau_{ij}(x_i,x_j)$ are either $x_i$ or $x_j$, i.e., one of the features pairs selected by the previous step. The resulting ranking models can achieve ranking performance close to unconstrained LambdaMART models without all their complexity, thus trading off between effectiveness and explainability.
    
\end{itemize}

\begin{table}[!ht]
\caption{Ablation Study Results. The table reports the P@1, NDCG@3, R@5, and MRR scores of \model and the baselines representing its different components. Scores marked with the $\blacktriangledown$ symbol indicate that \model statistically outperforms the corresponding baseline according to the paired t-test with Bonferroni correction and p-value<0.05. Conversely, the $\blacktriangle$ symbol indicates that the corresponding baseline statistically outperforms \model. Underlined values indicate that the baseline has a higher score than \model, but the difference is not statistically significant.}
\begin{adjustbox}{width=\textwidth}
\begin{tabular}{l|cccc|cccc|cccc}
\diagbox[width=\dimexpr \textwidth/8+2\tabcolsep\relax, height=1cm]{Model}{Dataset} & \multicolumn{4}{c|}{StackOverflow} & \multicolumn{4}{c|}{Unix} & \multicolumn{4}{c}{AskUbuntu} \\ \cline{1-13} 
                & P@1 & NDCG@3 & R@5 & MRR & P@1 & NDCG@3 & R@5 & MRR & P@1 & NDCG@3 & R@5 & MRR \\
$BC$            & 0.009$\blacktriangledown$ & 0.017$\blacktriangledown$ & 0.045$\blacktriangledown$ & 0.030$\blacktriangledown$ & 0.046$\blacktriangledown$ & 0.087$\blacktriangledown$ & 0.165$\blacktriangledown$ & 0.122$\blacktriangledown$ & 0.006$\blacktriangledown$ & 0.010$\blacktriangledown$ & 0.051$\blacktriangledown$& 0.044$\blacktriangledown$\\
$BM25$          & 0.209$\blacktriangledown$ & 0.259$\blacktriangledown$ & 0.325$\blacktriangledown$ & 0.259$\blacktriangledown$ & 0.116$\blacktriangledown$ & 0.217$\blacktriangledown$ & 0.396$\blacktriangledown$ & 0.254$\blacktriangledown$ & 0.098$\blacktriangledown$ & 0.196$\blacktriangledown$ & 0.352$\blacktriangledown$& 0.225$\blacktriangledown$\\
$\model_{NB}$     & 0.062$\blacktriangledown$ & 0.127$\blacktriangledown$ & 0.236$\blacktriangledown$ & 0.140$\blacktriangledown$ & 0.27$\blacktriangledown$ & 0.353$\blacktriangledown$ & 0.494$\blacktriangledown$ & 0.376$\blacktriangledown$ & 0.126$\blacktriangledown$ & 0.182$\blacktriangledown$ & 0.299$\blacktriangledown$& 0.208$\blacktriangledown$\\
$\model_{Lin}$    & 0.263$\blacktriangledown$ & 0.383$\blacktriangledown$ & 0.569$\blacktriangledown$ & 0.406$\blacktriangledown$ & 0.297$\blacktriangledown$ & 0.382$\blacktriangledown$ & 0.542$\blacktriangledown$ & 0.41$\blacktriangledown$ & 0.184$\blacktriangledown$ & 0.285$\blacktriangledown$ & 0.440$\blacktriangledown$& 0.307$\blacktriangledown$\\
$\model_{SL}$     & 0.431 & 0.565$\blacktriangledown$ & 0.750 & 0.564$\blacktriangledown$ & \underline{0.336} & \underline{0.430} & \underline{0.585} & \underline{0.452} & 0.255$\blacktriangle$ & 0.363$\blacktriangle$ & 0.513$\blacktriangle$ & 0.375$\blacktriangle$ \\
$\model_{CB}$     & 0.455 & 0.589 & \underline{0.761} & 0.587 & \underline{0.332} & 0.424 & 0.573$\blacktriangledown$ & 0.444$\blacktriangledown$ & \underline{0.244} & \underline{0.350} & \underline{0.502} & \underline{0.364} \\
$\model_{NoRW}$   & 0.446 & 0.582 & 0.744$\blacktriangledown$ & 0.573$\blacktriangledown$ & \underline{0.334} & 0.425 & 0.575 & 0.436$\blacktriangledown$ & \underline{0.249} & 0.342 & 0.472$\blacktriangledown$ & 0.345$\blacktriangledown$\\ 
$\model_{IlMart}$ & 0.360$\blacktriangledown$ & 0.496$\blacktriangledown$ & 0.676$\blacktriangledown$ & 0.502$\blacktriangledown$ & 0.324 & 0.416$\blacktriangledown$ & 0.565$\blacktriangledown$ & 0.438$\blacktriangledown$ & 0.203$\blacktriangledown$ & 0.305$\blacktriangledown$ & 0.468$\blacktriangledown$ & 0.327$\blacktriangledown$ \\\hline
\textbf{\model}   & 0.459 & 0.592 & 0.760 & 0.590 & 0.331 & 0.425 & 0.581 & 0.448  & 0.241 & 0.345 & 0.500 & 0.363 \\ \hline \hline
\diagbox[width=\dimexpr \textwidth/8+2\tabcolsep\relax, height=1cm]{Model}{Dataset} & \multicolumn{4}{c|}{Server Fault} & \multicolumn{4}{c|}{Physics} & \multicolumn{4}{c}{Mathematics} \\ \cline{1-13} 
                & P@1 & NDCG@3 & R@5     & MRR & P@1 & NDCG@3 & R@5     & MRR & P@1 & NDCG@3 & R@5  & MRR \\
$BC$            & 0.010$\blacktriangledown$ & 0.040$\blacktriangledown$ & 0.099$\blacktriangledown$ & 0.071$\blacktriangledown$ & 0.014$\blacktriangledown$ & 0.027$\blacktriangledown$ & 0.093$\blacktriangledown$ & 0.059$\blacktriangledown$ & 0.002$\blacktriangledown$ & 0.040$\blacktriangledown$ & 0.136$\blacktriangledown$ & 0.059$\blacktriangledown$\\
$BM25$          & 0.094$\blacktriangledown$ & 0.152$\blacktriangledown$ & 0.251$\blacktriangledown$ & 0.167$\blacktriangledown$ & 0.103$\blacktriangledown$ & 0.189$\blacktriangledown$ & 0.334$\blacktriangledown$ & 0.226$\blacktriangledown$ & 0.111$\blacktriangledown$ & 0.199$\blacktriangledown$ & 0.361$\blacktriangledown$ & 0.236$\blacktriangledown$\\
$\model_{NB}$     & 0.21$\blacktriangledown$ & 0.253$\blacktriangledown$ & 0.364$\blacktriangledown$ & 0.285$\blacktriangledown$ & 0.101$\blacktriangledown$ & 0.162$\blacktriangledown$ & 0.277$\blacktriangledown$ & 0.196$\blacktriangledown$ & 0.084$\blacktriangledown$ & 0.145$\blacktriangledown$ & 0.278$\blacktriangledown$ & 0.178$\blacktriangledown$\\
$\model_{LIN}$    & 0.260 & 0.368 & 0.524$\blacktriangledown$ & 0.388 & 0.139$\blacktriangledown$& 0.217$\blacktriangledown$ & 0.357$\blacktriangledown$ & 0.253$\blacktriangledown$ & 0.177$\blacktriangledown$ & 0.27$\blacktriangledown$ & 0.427$\blacktriangledown$ & 0.301$\blacktriangledown$\\
$\model_{SL}$     & \underline{0.279} & \underline{0.389} & 0.547 & \underline{0.406} & 0.220 & 0.299 & 0.440 & 0.331 & \underline{0.208} & 0.315 & 0.471 & 0.335 \\
$\model_{CB}$     & 0.267 & 0.377 & 0.538 & 0.395 & 0.215 & 0.294 & 0.429 & 0.325$\blacktriangledown$ & 0.219$\blacktriangle$ & 0.327$\blacktriangle$ & \underline{0.482} & \underline{0.345} \\
$\model_{NORW}$   & \underline{0.272} & 0.378 & 0.527$\blacktriangledown$ & 0.382$\blacktriangledown$ & 0.218 & 0.295 & 0.429 & 0.313$\blacktriangledown$ & \underline{0.210} & 0.313 & 0.469 & 0.321$\blacktriangledown$\\ 
$\model_{IlMart}$  & 0.250 & 0.364$\blacktriangledown$ & 0.529$\blacktriangledown$ & 0.384$\blacktriangledown$ & 0.195$\blacktriangledown$ & 0.276$\blacktriangledown$ & 0.418$\blacktriangledown$ & 0.309$\blacktriangledown$  & 0.197$\blacktriangledown$ & 0.301$\blacktriangledown$ & 0.475 & 0.329$\blacktriangledown$ \\ \hline
\textbf{\model}   & 0.266 & 0.382 & 0.547 & 0.400 & 0.221 & 0.300 & 0.440 & 0.332 & 0.207 & 0.316 & 0.478 & 0.340 \\ \hline
\end{tabular}
\end{adjustbox}
\label{tab:ablation_results}
\end{table}

\paragraph{\textbf{Discussion}}
Table \ref{tab:ablation_results} reports the results of the ablation study. We mark statistically significant performance gains/losses with respect to \model with the symbols $\blacktriangledown$ and $\blacktriangle$ (paired t-test with p-value<0.05 and Bonferroni correction), and we underline the performance figures numerically greater than those of \model.

\model consistently outperforms in all examined communities some baseline models, including $BC$, $BM25$, $\model_{NB}$, and $\model_{Lin}$. A recurring pattern emerges across communities, revealing $BC$ and $\model_{NB}$ as the least effective baselines, both relying exclusively on network aspects. In contrast, $BM25$'s performance highlights the significance of the content-based component, as evidenced by a minimum of 10\% of queries across communities being best addressed by experts who have previously answered similar questions. $\model_{Lin}$ consistently exhibits lower performance compared to \model, underscoring the efficacy of \model's LtR methodology in capturing hidden relationships among expert features.

$\model_{SL}$ shows slightly higher or equal performance across communities except for StackOverflow and AskUbuntu. In the case of StackOverflow, \model outperforms a statistically significant margin $\model_{SL}$, while on AskUbuntu, the opposite holds, hinting at community-specific influences. This observation aligns with the Silhouette values described in Table \ref{tab:stats}, which quantify the clustering effectiveness of layers within the MLG. StackOverflow's superior Silhouette value of 0.805 indicates a robust subdivision of tags into clusters, benefiting \model's exploitation of topic layers. Conversely, AskUbuntu, with the lowest Silhouette value of 0.391, experiences challenges, contributing to $\model_{SL}$'s higher statistical performance. Consequently, the MLG proves beneficial when distinct community topics are evident. Moreover, the values of $\model_{CB}$ underscore content supremacy over the network component, although incorporating both components in \model marginally enhances system performances for most communities. 
Finally, $\model_{NoRW}$ exhibits lower recall (R@5) without a decrease in precision (P@1). The similar P@1 scores between \model and $\model_{NoRW}$ result from \model's random walk mechanism, which identifies a larger set of relevant experts. 
\model significantly improves recall (R@5) by consistently including the correct expert within the top five positions. However, this larger candidate set makes it more challenging to rank the best expert in the first position, thus keeping P@1 similar for both models.
Overall, \model's random walk enhances recall by expanding the candidate set, even though it complicates achieving the highest precision score.

\begin{figure}
    \centering
    \includegraphics[width=0.9\linewidth]{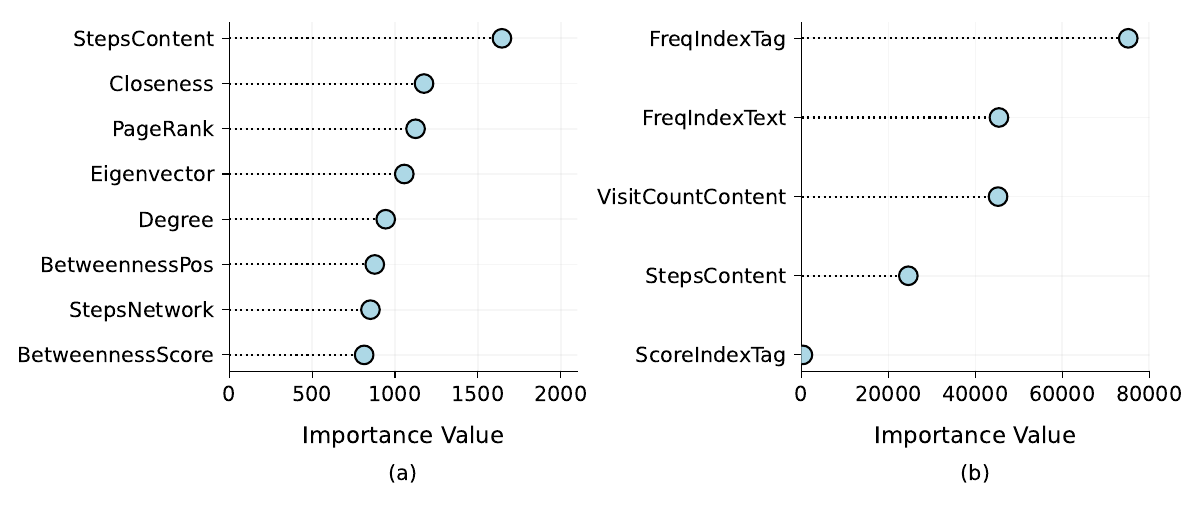}
    \caption{LtR Algorithm's Feature Importances. Figure (a) on the left displays the eight most important features for the \model LtR algorithm. Figure (b) illustrates the feature importance values for all features considered by \model\textsubscript{IlMart}.}
    \label{fig:feat_imp}
\end{figure}

\begin{figure}[!ht]
    \centering    \includegraphics[width=0.95\textwidth]{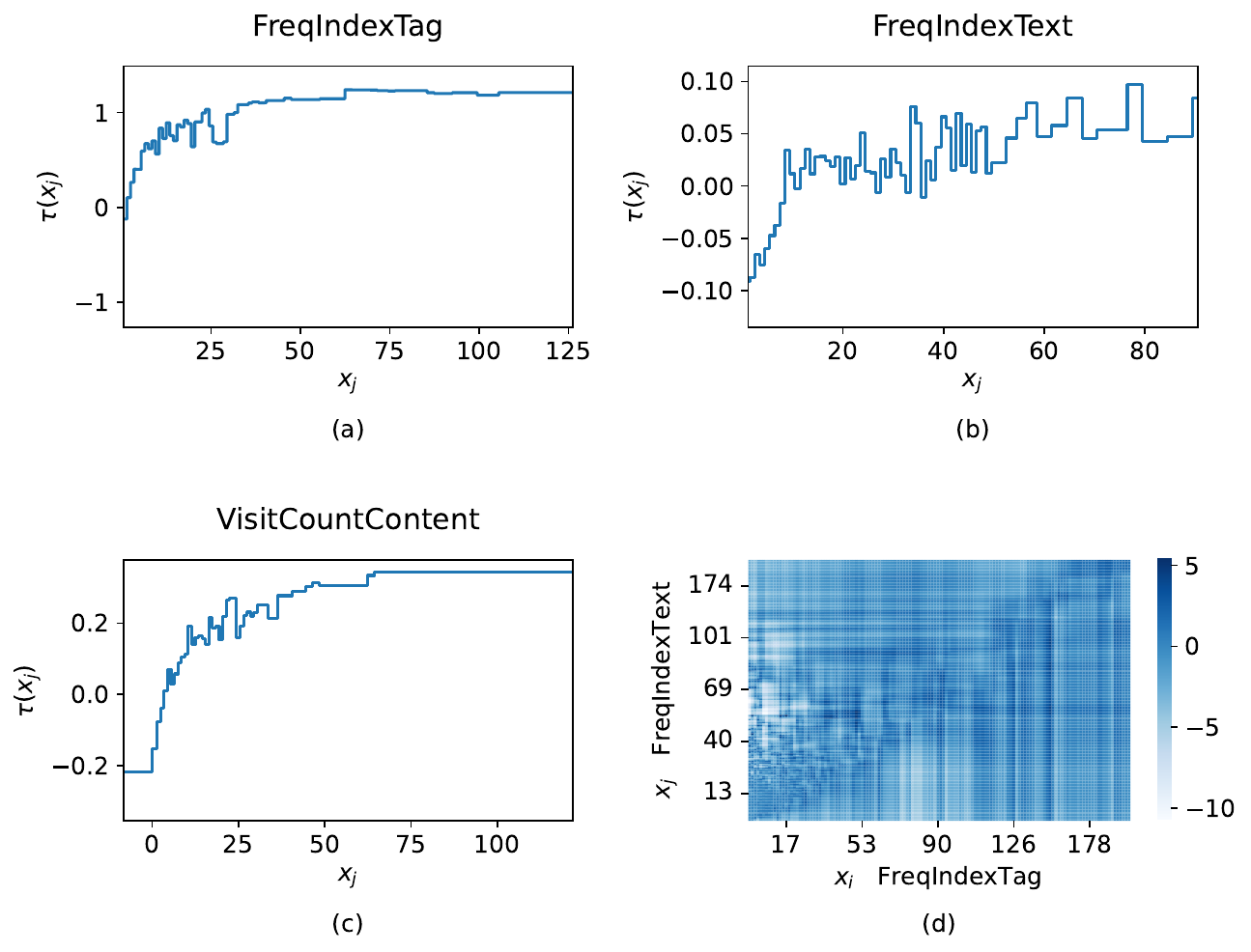}
    \caption{The three most important main and interaction effects of \model\textsubscript{IlMart} on the StackOverflow dataset. Figures (a), (b), and (c) show the main effects of FreqIndexTag, FreqIndexText, and VisitCountContent, respectively. The x-axis represents the values the feature can have, while the y-axis represents the corresponding contribution of the feature to the predicted final score. Figure (d) illustrates the most important interaction effect learned by \model\textsubscript{IlMart}, composed of the features FreqIndexTag (x-axis) and FreqIndexText (y-axis). The color bar indicates the interaction contribution.}
    \label{fig:ilmart_inter}
\end{figure}

\paragraph{\textbf{Interpretability}} $\model_{IlMart}$  consistently exhibits statistically lower performance than \model, with the most notable disparity occurring in the StackOverflow community, where the difference reaches 27.5\% for P@1 and 12.43\% for R@5. In other communities, the variations are still statistically significant, ranging from a maximum of 18.71\% (AskUbuntu) to a minimum of 2.16\% (Unix) in terms of P@1. However, the differences diminish significantly when considering R@5, with the highest value being 6.83\% (AskUbuntu) and no statistical differences in Mathematics, indicating $\model_{IlMart}$ proficiency in placing the top respondent within the first five positions.

$\model_{IlMart}$ capacity to offer model result explanations may offset the observed performance decline. \model, on the other hand, employs a more sophisticated decision tree-based LtR approach, which contributes to higher performance but offers only minimal clarity regarding the significance of different factors influencing the ranking result. Figure \ref{fig:feat_imp} illustrates the feature importance of both models (LambdaMart and $\model_{IlMart}$) for the StackOverflow community. It reports LambdaMart's top height features and all those considered by $\model_{IlMart}$, all of which are present among LambdaMart's top eight features, albeit in a different order. For example, the ScoreIndexTag feature, most used by LambdaMart, is the least used by $\model_{IlMart}$.
Most of the crucial features for LambdaMart are associated with the content-based method. Nevertheless, features from the network-based method, such as Closeness, PageRank, and AvgWeight, exhibit high importance values. On the other hand, all the features selected by $\model_{IlMart}$ are content-based, highlighting the importance of the text component for the EF task.

Figure \ref{fig:ilmart_inter} illustrates how $\model_{IlMart}$ can be used for interpretability purposes, showcasing the contribution to the predicted score of the three IlMart main effects with the highest feature importance values (i.e., FreqIndexTag, VisitCountContent, and FreqIndexText) and the most important interaction effect learned from $\model_{IlMart}$ for the StackOverflow community. High values of these features indicate proficient expert competence. 
The 2D plots and the heatmap are built by aggregating all the trees using the same features, i.e., by representing the contribution of each main effect $\tau(x_j)$ and each interaction effect $\tau(x_i, x_j)$, where $\tau(x_j)$ and $\tau_{ij}(x_i,x_j)$ are ensembles of trees.
Through Figure \ref{fig:ilmart_inter}, we can analyze instances where \model successfully places the best answerer in the first position and cases where it fails. Table \ref{tab:test_data} presents four question examples: the first two showcase instances where \model successfully identifies the best expert by placing them in the first position. For a better understanding, we report the features of the top two ranked experts. Conversely, for the last two examples where \model fails, we show the feature values of the experts in ranking order up to the position where the ranker placed the best expert. The \textit{Rank} column indicates the expert's ranking position in the final list. In contrast, the \textit{BestExpert} column serves as a label, with a value equal to 1 for the ground truth expert and 0 otherwise.

A notable pattern emerges in both instances of \model's success and failure: the expert occupying the first position consistently demonstrates higher values for the features evaluated by the $\model_{IlMart}$ ranker. This results in a greater contribution and, consequently, a higher final score, as shown in Figure \ref{fig:ilmart_inter}. 
For example, consider the first instance in Table \ref{tab:test_data}, with $QID=65438592$, and focus on the feature \textit{FreqIndexTag}. The BestExpert (first row of Table \ref{tab:test_data}) has a value of 45, corresponding to a contribution of approximately 1, as shown in the upper left figure. In contrast, the expert placed in the second position (second row of Table \ref{tab:test_data}) has a value of 7, contributing to a measure of less than 0.5 (half of the first). By combining all feature contributions computed with the aid of Figure \ref{fig:ilmart_inter}, we can comprehend the final ranking and understand why the model could or could not place the BestExpert in the first position.

\begin{table}[]
\caption{Examples of \model\textsubscript{IlMart} Decision-Making Process. The table presents four examples of questions: two where \model\textsubscript{IlMart} successfully identified the right expert (i.e., placing the right expert in the first position of the final ranking) and two where it was unsuccessful. For each question, the table reports the values of FreqIndexTag, VisitCountContent, and FreqIndexText, along with the final expert Score, the position in the final ranking, and the BestExpert label.}
\centering
\begin{tabular}{@{}ccccccc@{}}
\toprule
QID & \multicolumn{1}{l}{FreqIndexTag} & \multicolumn{1}{l}{VisitCountContent} & \multicolumn{1}{l}{FreqIndexText} & \multicolumn{1}{l}{Score} & \multicolumn{1}{l}{Rank} & \multicolumn{1}{l}{BestExpert} \\ \midrule
\multirow{2}{*}{65438592} & 45 & 17 & 54 & 2.39 & 1 & 1 \\
 & 7 & 7 & 7 & 1.96 & 2 & 0 \\ \midrule
\multirow{2}{*}{65438859} & 2 & 10 & 13 & -0.12 & 1 & 1 \\
 & 1 & 1 & 9 & -1.7 & 2 & 0 \\ \midrule
\multirow{2}{*}{65438631} & 81 & 62 & 82 & 3.29 & 1 & 0 \\
 & 30 & 25 & 30 & 2.83 & 2 & 1 \\ \midrule
\multirow{3}{*}{65439018} & 229 & 14 & 202 & 3.5 & 1 & 0 \\
 & 14 & 5 & 12 & 2.58 & 2 & 0 \\
 & 11 & 9 & 16 & 2.45 & 3 & 1 \\ \bottomrule
\end{tabular}
\label{tab:test_data}
\end{table}

\subsection{State-of-the-art Baselines}
\label{sec:baselines}

In this section, we evaluate \model based on the two approaches for the EF task discussed in Section \ref{sec:related}. As done in Section \ref{sec:ablation}, to assess the statistical significance of the performance gains/losses ($\blacktriangle$ and $\blacktriangledown$) measured with our experiments, we conducted a paired t-test with a p-value threshold of 0.05 and Bonferroni correction. Moreover, we underline the performance numerically (not statistically) greater than those of \model.

\paragraph{\textbf{Expert Ranking.}} 

\begin{table}
\caption{Expert Ranking Comparison. The table reports the P@1, NDCG@3, R@5, and MRR scores of \model and the baselines in the Expert Ranking category for all selected StackExchange communities. Scores marked with the $\blacktriangledown$ symbol indicate that \model statistically outperforms the corresponding baseline according to the paired t-test with Bonferroni correction and p-value<0.05. Conversely, the $\blacktriangle$ symbol indicates that the corresponding baseline statistically outperforms \model. Underlined values indicate that the baseline has a higher score than \model, but the difference is not statistically significant.}

\begin{adjustbox}{width=\textwidth}
\begin{tabular}{l|cccc|cccc|cccc}
\diagbox{Model}{Dataset} & \multicolumn{4}{c|}{Stack Overflow} & \multicolumn{4}{c|}{Unix} & \multicolumn{4}{c}{Ask Ubuntu} \\ \cline{1-13} 
& P@1 & NDCG@3 & R@5 & MRR & P@1& NDCG@3 &  R@5 & MRR & P@1 & NDCG@3 & R@5 & MRR \\
BM25 & 0.257$\blacktriangledown$ & 0.366$\blacktriangledown$ & 0.513$\blacktriangledown$ & 0.381$\blacktriangledown$ & 0.253$\blacktriangledown$ & 0.348$\blacktriangledown$ & 0.492$\blacktriangledown$ & 0.365$\blacktriangledown$ & 0.141$\blacktriangledown$ & 0.237$\blacktriangledown$ & 0.389$\blacktriangledown$ & 0.261$\blacktriangledown$ \\
BM25+TAG & 0.307$\blacktriangledown$ & 0.431$\blacktriangledown$ & 0.604$\blacktriangledown$ & 0.441$\blacktriangledown$ & 0.260$\blacktriangledown$ & 0.345$\blacktriangledown$ & 0.494$\blacktriangledown$ & 0.368$\blacktriangledown$ & 0.151$\blacktriangledown$ & 0.256$\blacktriangledown$ & 0.415$\blacktriangledown$ & 0.277$\blacktriangledown$ \\
MiniLM & 0.282$\blacktriangledown$ & 0.384$\blacktriangledown$ & 0.527$\blacktriangledown$ & 0.399$\blacktriangledown$ & 0.262$\blacktriangledown$ & 0.352$\blacktriangledown$ & 0.485$\blacktriangledown$ & 0.367$\blacktriangledown$ & 0.158$\blacktriangledown$ & 0.259$\blacktriangledown$ & 0.412$\blacktriangledown$ & 0.279$\blacktriangledown$ \\
MiniLM+TAG & 0.320$\blacktriangledown$ & 0.441$\blacktriangledown$ & 0.624$\blacktriangledown$ & 0.456$\blacktriangledown$ & 0.269$\blacktriangledown$ & 0.358$\blacktriangledown$ & 0.492$\blacktriangledown$ & 0.371$\blacktriangledown$ & 0.172$\blacktriangledown$ & 0.273$\blacktriangledown$ & 0.419$\blacktriangledown$ & 0.292$\blacktriangledown$ \\
DistilBERT & 0.288$\blacktriangledown$ & 0.394$\blacktriangledown$ & 0.542$\blacktriangledown$ & 0.407$\blacktriangledown$ & 0.268$\blacktriangledown$ & 0.356$\blacktriangledown$ & 0.492$\blacktriangledown$ & 0.373$\blacktriangledown$ & 0.161$\blacktriangledown$ & 0.264$\blacktriangledown$ & 0.415$\blacktriangledown$ & 0.283$\blacktriangledown$ \\
DistilBERT+TAG & 0.337$\blacktriangledown$ & 0.455$\blacktriangledown$ & 0.627$\blacktriangledown$ & 0.468$\blacktriangledown$ & 0.276$\blacktriangledown$ & 0.363$\blacktriangledown$ & 0.492$\blacktriangledown$ & 0.376$\blacktriangledown$ & 0.178$\blacktriangledown$ & 0.282$\blacktriangledown$ & 0.438$\blacktriangledown$ & 0.302$\blacktriangledown$ \\
t-BGER & 0.254$\blacktriangledown$ & 0.338$\blacktriangledown$ & 0.478$\blacktriangledown$ & 0.36$\blacktriangledown$ & 0.25$\blacktriangledown$ & 0.314$\blacktriangledown$ & 0.43$\blacktriangledown$ & 0.345$\blacktriangledown$ & 0.202$\blacktriangledown$ & 0.312$\blacktriangledown$ & \underline{0.508} & 0.335$\blacktriangledown$ \\ \hline
\textbf{\model}   & 0.459 & 0.592 & 0.760 & 0.590 & 0.331 & 0.425 & 0.581 & 0.448  & 0.241 & 0.345 & 0.500 & 0.363 \\ \hline
\diagbox{Model}{Dataset} & \multicolumn{4}{c|}{Server Fault} & \multicolumn{4}{c|}{Physics} & \multicolumn{4}{c}{Mathematics} \\ \cline{1-13} 
 & P@1 & NDCG@3 & R@5 & MRR & P@1 & NDCG@3 & R@5 & MRR & P@1  & NDCG@3 & R@5 & MRR \\
BM25 & 0.199$\blacktriangledown$ & 0.286$\blacktriangledown$ & 0.430$\blacktriangledown$ & 0.311$\blacktriangledown$ & 0.164$\blacktriangledown$ & 0.234$\blacktriangledown$ & 0.364$\blacktriangledown$ & 0.267$\blacktriangledown$ & 0.149$\blacktriangledown$ & 0.235$\blacktriangledown$ & 0.378$\blacktriangledown$ & 0.265$\blacktriangledown$ \\
BM25+TAG & 0.218$\blacktriangledown$ & 0.309$\blacktriangledown$ & 0.475$\blacktriangledown$ & 0.331$\blacktriangledown$ & 0.166$\blacktriangledown$ & 0.239$\blacktriangledown$ & 0.374$\blacktriangledown$ & 0.271$\blacktriangledown$ & 0.149$\blacktriangledown$ & 0.232$\blacktriangledown$ & 0.377$\blacktriangledown$ & 0.258$\blacktriangledown$ \\
MiniLM & 0.200$\blacktriangledown$ & 0.293$\blacktriangledown$ & 0.436$\blacktriangledown$ & 0.313$\blacktriangledown$ & 0.172$\blacktriangledown$ & 0.248$\blacktriangledown$ & 0.382$\blacktriangledown$ & 0.278$\blacktriangledown$ & 0.159$\blacktriangledown$ & 0.248$\blacktriangledown$ & 0.403$\blacktriangledown$ & 0.277$\blacktriangledown$ \\
MiniLM+TAG & 0.224$\blacktriangledown$ & 0.315$\blacktriangledown$ & 0.476$\blacktriangledown$ & 0.337$\blacktriangledown$ & 0.172$\blacktriangledown$ & 0.25$\blacktriangledown$ & 0.386$\blacktriangledown$ & 0.279$\blacktriangledown$ & 0.161$\blacktriangledown$ & 0.248$\blacktriangledown$ & 0.405$\blacktriangledown$ & 0.277$\blacktriangledown$ \\
DistilBERT & 0.215$\blacktriangledown$ & 0.302$\blacktriangledown$ & 0.449$\blacktriangledown$ & 0.325$\blacktriangledown$ & 0.178$\blacktriangledown$ & 0.256$\blacktriangledown$ & 0.386$\blacktriangledown$ & 0.284$\blacktriangledown$ & 0.170$\blacktriangledown$ & 0.263$\blacktriangledown$ & 0.411$\blacktriangledown$ & 0.291$\blacktriangledown$ \\
DistilBERT+TAG & 0.223$\blacktriangledown$ & 0.319$\blacktriangledown$ & 0.478$\blacktriangledown$ & 0.341$\blacktriangledown$ & 0.179$\blacktriangledown$ & 0.254$\blacktriangledown$ & 0.384$\blacktriangledown$ & 0.283$\blacktriangledown$ & 0.17$\blacktriangledown$ & 0.262$\blacktriangledown$ & 0.414$\blacktriangledown$ & 0.291$\blacktriangledown$ \\
t-BGER & 0.275$\blacktriangle$ & 0.356$\blacktriangledown$ & 0.474$\blacktriangledown$ & 0.375$\blacktriangledown$ & 0.080$\blacktriangledown$ & 0.151 $\blacktriangledown$& 0.258$\blacktriangledown$ & 0.180$\blacktriangledown$ & 0.122$\blacktriangledown$ & 0.169$\blacktriangledown$ & 0.295$\blacktriangledown$ & 0.215$\blacktriangledown$ \\ \hline
\textbf{\model}   & 0.266 & 0.382 & 0.547 & 0.400 & 0.221 & 0.300 & 0.440 & 0.332 & 0.207 & 0.316 & 0.478 & 0.340 \\ \hline
\end{tabular}
\end{adjustbox}

\label{tab:tuef_config}
\end{table}

The first set of baselines follows the \textit{Expert Ranking} configuration, which involves an end-to-end process of selecting users as experts, computing the similarity between a candidate expert and the query, and subsequently ranking the candidate experts. \model adopts this configuration, and we compare it with the models introduced in \cite{kasela2023se}, enabling us to evaluate the effectiveness of current neural re-rankers on this task, as well as the \textit{t-BGER} model \cite{krishna2023temporal}.

Kasela et al. \cite{kasela2023se} employ a two-stage ranking architecture, where the initial stage uses a recall-oriented retriever, and the subsequent stage utilizes a precision-oriented approach. In the first stage, they use Elastic Search's \textit{BM25} model to generate an ordered and limited list of candidate experts. In the second stage, they apply a linear combination of scores to re-rank these experts. This combination includes (i) the BM25 score, (ii) the similarity score computed by neural network-based re-ranker approaches, and for three of the six proposed baselines, (iii) the score generated by a personalized model based on user history.
The neural network-based re-ranker models are $DistilBERT$\footnote{\url{https://huggingface.co/distilbert-base-uncased}} and $MiniLM$\footnote{\url{https://huggingface.co/sentence-transformers/all-MiniLM-L6-v2}}.
Both pre-trained models are fine-tuned for ten epochs for each community considered. The personalized model, called \textit{TAG}, captures the similarity between the topics previously addressed by the asker and the answerer based on the tags in the questions asked and the answers provided. 
Finally, the scores are combined and computed as the weighted sum of the normalized model scores. The models' weights are determined using a grid search on the validation set. The proposed baselines are \textit{BM25}, \textit{BM25+TAG}, \textit{MiniLM}, \textit{MiniLM+TAG}, \textit{DistilBERT}, and \textit{DistilBERT+TAG}. Although not explicitly stated for readability, the last four baselines also incorporate \textit{BM25}.

In contrast, Krishna et al. \cite{krishna2023temporal} introduce \textit{t-BGER}, a temporal weighted bipartite graph model. Their study emphasizes the importance of utilizing tags for expert identification while reducing task complexity. The model is based on a user-tag bipartite graph, where the edge weight reflects the user's activity related to the specific tag. Additionally, they incorporate temporal discounting into this graph, assigning higher weights to recent activities. Finally, they apply resource allocation techniques through network-based inference on the bipartite graph, effectively managing highly sparse data.

As reported in Table \ref{tab:tuef_config}, among the baselines, \textit{DistilBERT+TAG} shows superior performance. However, \textit{t-BGER} demonstrates better scores for the AskUbuntu and Server Fault communities. Overall, \model surpasses all baselines across various metrics, except for P@1 in the Server Fault community, where \textit{t-BGER} exhibits a relative improvement of 3.38\%, and R@5 in the AskUbuntu community, where \textit{t-BGER} has a slightly higher score. Excluding these two exceptions, \model records minimum relative improvements of 19.31\% for P@1 in the AskUbuntu community, 7.30\% for NDCG@3, 15.40\% for R@5, and 6.67\% for MRR in the Server Fault community. In the remaining communities, \model reports higher relative improvements across all other metrics.

\paragraph{\textbf{Expert Subsample Ranking.}} The second set of baselines follows the Expert Ranking evaluation approach and aims to rank a small set of 20 experts when presented with a new query (see Section \ref{sec:related}). This set always includes (i) the users who have provided the answers, including the best answerer, and (ii) a variable number of users selected to reach a total of 20 candidate experts, chosen from the top 10\% of users identified as the most active answerers.
Since this definition of the EF task deviates from the  task previously considered by requiring the prior knowledge of who will answer the query, we adapted \model to allow a fair comparison with the following two state-of-the-art models for the EF task, whose code is publicly available:
\begin{itemize}
    \item \textbf{NeRank \cite{li2019personalized}}: It models the CQA platform as a heterogeneous network to learn representations for question raisers and question answerers through a metapath-based algorithm. Using this heterogeneous network, NeRank preserves relationship information while modeling the question's content with a single-layer LSTM. Finally, a CNN assigns a score to each expert for a given question, representing the probability of the expert providing the accepted answer.
    \item \textbf{PMEF \cite{peng2022towards}}: It consists of three main modules: a multi-view question encoder for learning comprehensive question features, an intra-view encoder to discover view-specific interactions among experts and target questions, and an inter-view encoder designed to extract expert/question features by integrating different view information in a personalized manner.
\end{itemize}

We always consider the questions selected for the experimentation phase for each community. However, the test set is smaller due to constraints imposed by NeRank, PMEF, and, in this specific setup, \model as well. Specifically, both NeRank and PMEF models require that the user posing the question has previously asked other questions to capture the interactions between the questioner and responder. Additionally, the test set includes only those questions for which \model, through graph exploration, included the best answerer in the set of potential experts.
Moreover, to ensure a fair comparison among the three models, we required each to rank the same set of experts. This set comprises users who have answered the question, including the best answerer, and additional experts randomly chosen from those identified by \model as candidates. This procedure is the fairest, considering the way \model selects candidate experts. Specifically, as a topic-based model, \model chooses hard-negative samples because they are always users who have previously answered precisely the topics related to the query. In contrast, following the procedure of NeRank and PMEF and choosing from the top 10\% of the community's best answerers would lead to selecting users who could be irrelevant to the query topics, thus making it easier to rank them lower.
We consider in the comparison also \model and $\model_{IlMart}$ to study the behavior of the interpretable version of \model under this different configuration.

\begin{table}[!ht]
\caption{Expert Subsample Ranking Comparison. The table reports the P@1, NDCG@3, R@5, and MRR scores of \model and the baselines in the Expert Ranking category for all selected StackExchange communities. Scores marked with the $\blacktriangledown$ symbol indicate that \model statistically outperforms the corresponding baseline according to the paired t-test with Bonferroni correction and p-value<0.05. Conversely, the $\blacktriangle$ symbol indicates that the corresponding baseline statistically outperforms \model. Underlined values indicate that the baseline has a higher score than \model, but the difference is not statistically significant.}

\begin{adjustbox}{width=\textwidth}
\begin{tabular}{l|cccc|cccc|cccc}
\diagbox{Model}{Dataset} & \multicolumn{4}{c|}{StackOverflow} & \multicolumn{4}{c|}{Unix} & \multicolumn{4}{c}{Ask Ubuntu} \\ \cline{1-13} 
& P@1 & NDCG@3 & R@5 & MRR & P@1 & NDCG@3 & R@5 & MRR & P@1 & NDCG@3 & R@5 & MRR\\
Nerank & 0.313$\blacktriangledown$ & 0.460$\blacktriangledown$ & 0.718$\blacktriangledown$ & 0.491$\blacktriangledown$ & 0.385$\blacktriangledown$ & 0.490$\blacktriangledown$ & 0.646$\blacktriangledown$ & 0.514$\blacktriangledown$ & 0.217$\blacktriangledown$ & 0.364$\blacktriangledown$ & 0.651$\blacktriangledown$ & 0.406$\blacktriangledown$ \\
PMEF & 0.116$\blacktriangledown$ & 0.213$\blacktriangledown$ & 0.422$\blacktriangledown$ & 0.273$\blacktriangledown$ & 0.429$\blacktriangledown$ & 0.556$\blacktriangledown$ & 0.718$\blacktriangledown$ & 0.567$\blacktriangledown$ & 0.298$\blacktriangledown$ & 0.472$\blacktriangledown$ & 0.740$\blacktriangledown$ & 0.492$\blacktriangledown$ \\
$\model_{IlMart}$ & 0.700$\blacktriangledown$ & 0.815$\blacktriangledown$ & 0.939$\blacktriangledown$ & 0.804$\blacktriangledown$ & \underline{0.619} & 0.744$\blacktriangle$ & 0.899$\blacktriangledown$ & \underline{0.741} & 0.528$\blacktriangledown$ & 0.673$\blacktriangledown$ & 0.875$\blacktriangledown$ & 0.673$\blacktriangledown$ \\\hline
\textbf{\model} & 0.800 & 0.883 & 0.989 & 0.877 & 0.611 & 0.738 & 0.906 & 0.736 & 0.574 & 0.718 & 0.901 & 0.712 \\ \hline
\diagbox{Model}{Dataset} & \multicolumn{4}{c|}{Server Fault} & \multicolumn{4}{c|}{Physics} & \multicolumn{4}{c}{Mathematics} \\ \cline{1-13} 
& P@1 & NDCG@3 & R@5 & MRR & P@1 & NDCG@3 & R@5 & MRR & P@1 & NDCG@3 & R@5 & MRR\\
Nerank & 0.292$\blacktriangledown$ & 0.403$\blacktriangledown$ & 0.597$\blacktriangledown$ & 0.442$\blacktriangledown$ & 0.187$\blacktriangledown$ & 0.341$\blacktriangledown$ & 0.621$\blacktriangledown$ & 0.380$\blacktriangledown$ & 0.233$\blacktriangledown$ & 0.372$\blacktriangledown$ & 0.608$\blacktriangledown$ & 0.407$\blacktriangledown$ \\
PMEF &  0.319$\blacktriangledown$ & 0.447$\blacktriangledown$ & 0.654$\blacktriangledown$ &  0.471$\blacktriangledown$ & 0.214$\blacktriangledown$ & 0.320$\blacktriangledown$ & 0.556$\blacktriangledown$ & 0.378$\blacktriangledown$ & 0.130$\blacktriangledown$ & 0.270$\blacktriangledown$ & 0.523$\blacktriangledown$ & 0.313$\blacktriangledown$ \\
$\model_{IlMart}$ & 0.498$\blacktriangledown$ & 0.668$\blacktriangledown$ & 0.850$\blacktriangledown$ & 0.660$\blacktriangledown$ & 0.520$\blacktriangle$ & 0.674$\blacktriangle$ & \underline{0.860} & 0.673$\blacktriangle$ & 0.507$\blacktriangledown$ & 0.627$\blacktriangledown$ & 0.805$\blacktriangledown$ & 0.642$\blacktriangledown$ \\\hline
\textbf{\model} & 0.594 & 0.745 & 0.916 & 0.733 & 0.487 & 0.635 & 0.858 & 0.640 & 0.648 & 0.669 & 0.907 & 0.760 \\ \hline
\end{tabular}
\end{adjustbox}
\label{tab:20config}
\end{table}

Table \ref{tab:20config} shows the results of the three models across different communities. PMEF consistently outperforms NeRank, except in the StackOverflow and Mathematics communities. However, \model exhibits significantly better performance than PMEF, with the smallest relative improvement being 42.42\% for P@1 and 29.8\% for MRR in the Unix community, and 21.92\% for R@5 in the AskUbuntu community. Other metrics in the remaining communities show even more substantial relative improvements than those mentioned.

$\model_{IlMart}$ consistently outperforms NeRank and PMEF. However, it exhibits lower performance across various communities compared to \model, except in the Unix and Physics communities, where it demonstrates statistical superiority in specific metrics. $\model_{IlMart}$ has already demonstrated performance comparable to \model in these communities during the ablation study (Table \ref{tab:ablation_results}). Consequently, $\model_{IlMart}$ is a promising compromise between interpretability and system accuracy for less \textit{topic-sensitive} communities.

Finally, note that \model performs considerably better in this set of experiments compared to the cases analyzed in Tables \ref{tab:ablation_results} and \ref{tab:tuef_config}. This improvement is the result of two factors: (i) the test set includes only queries for which \model successfully identified the best answerer during graph exploration, and (ii) the set of candidates to be ranked is limited to twenty users, approximately one-fifth of those that \model typically selects and sorts on average (see Table \ref{tab:stats}). We remark that this configuration and the specific settings for these experiments were required to ensure a fair comparison with the other considered algorithms.

\subsection{TUEF scalability}
\label{sec:scalability}

\begin{table}[h!]
\centering
\caption{Statistics for four datasets corresponding to 1, 2, 3, and 4 months of StackOverflow data, including the number of questions used for train and LtR, the number of answers, the number of users labeled as experts, the minimum number of accepted answers the experts have (MinAccAns), the number of tags associated with questions, the number of MLG’s layers, and the average length of experts lists to rank at inference time (AvgLists).}
\label{tab:tuef_so_statistics}
\begin{tabular}{ccccccccc}
\toprule
Months & Train & Answers & LtR & Experts & MinAccAns & Tags & Layers & AvgLists \\ \hline
1 & 38616 & 53717 & 8008 & 350 & 11 & 5365 & 10 & 111.03 \\ \hline
2 & 79451 & 110169 & 19991 & 577 & 14 & 7546 & 10 & 146.74 \\ \hline
3 & 122612 & 170131 & 33164 & 819 & 15 & 9136 & 10 & 182.68 \\ \hline
4 & 165320 & 230246 & 44821 & 1015 & 16 & 10514 & 10 & 186.06 \\ \bottomrule
\end{tabular}
\end{table}

\begin{figure}[h!]
    \centering
    \includegraphics[width=1\linewidth]{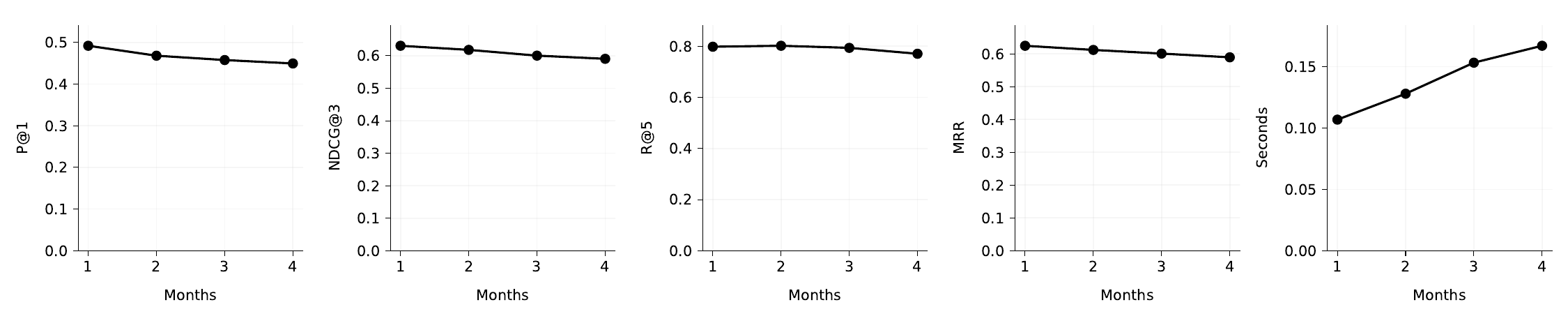}
    \caption{\model performance with four datasets corresponding to 1, 2, 3, and 4 months of StackOverflow data. The x-axis specifies the number of months considered, while the y-axis reports the performance metrics, including  P@1, NDCG@3, R@5, MRR computed on the same test set of 1,342 queries. The rightmost plot indicates the per-query  \model average inference time in seconds.}
    \label{fig:tuef_so_performance}
\end{figure}

Table \ref{tab:tuef_so_statistics} and Figure \ref{fig:tuef_so_performance} illustrate \model ability to scale and adapt to larger datasets while maintaining its effectiveness. The table shows key statistics for four datasets corresponding to 1, 2, 3, and 4 months of StackOverflow data, including the number of questions used for training the MLG and the LtR model, the number of answers, the number of users labeled as experts, the minimum number of accepted answers the experts have (MinAccAns), the number of tags associated with questions, the number of MLG’s Layers, and the average length of experts lists to rank at inference time (AvgLists).
The training set size increases approximately linearly, with a consistent monthly growth rate, expanding from 38,616 questions in the 1-month dataset to 165,320 questions in the 4-month dataset. The increase in the number of questions is associated with an increase in the number of unique tags, which rise from 5,365 to 10,514, indicating a richer set of topics being covered over time. 
As expected, using more data increases the number of identified experts (from 350 to 1,015), thereby enriching the expert base. This increase in identified experts by \model leads to a corresponding increment in the number of retrieved experts at inference time, as reported in the AvgList column.

Despite the significant growth in data volume, \model maintains noteworthy performance on a test set of 1,432 questions, as evidenced by the performance metrics provided (Figure \ref{fig:tuef_so_performance}). The P@1 metric slightly decreases from 0.492 (1-month) to 0.449 (4-month). The NDCG@3 metric decreases from 0.631 to 0.590, while the MRR slightly drops from 0.625 to 0.589. This decline is relatively minor, considering the significant increase in the training data and the set of experts.
However, the R@5 remains more stable for the datasets corresponding to 1, 2, and 3 months, and only slightly decreases to 0.771 for the 4-month dataset, indicating that the model consistently ranks the best answerer within the top five positions. The stability of the R@5 metric is particularly noteworthy, as it suggests that while \model may slightly reduce its performance in identifying the best experts at the top position, it still effectively includes them within the first 5 positions. 
One explanation for the observed slight decrease in performance in metrics other than R@5 is the increased difficulty the LtR algorithm faces when ranking a larger number of experts, as evidenced by the rise in the average length of lists of candidate experts (111 to 186). As the dataset grows, the expanding pool of candidate experts makes it more challenging to distinguish the best expert, leading to minor drops in Precision, NDCG, and MRR metrics. 
Finally, the rightmost plot shows the average latency of a single query, which increases from 0.107 seconds for the 1-month dataset to 0.167 seconds for the 4-month dataset. This increment can be attributed to the larger size of the CQA community in the 4-month dataset. Specifically, the greater number of questions leads to a larger and more dense MLG as more users and their interactions are included. Consequently, the RW applied to this denser MLG may perform more steps, resulting in the selection of more candidate experts to rank, thereby increasing the time required to perform the inference.
Overall, \model demonstrates robust scalability, effectively managing larger datasets while maintaining a high level of performance, with slight declines in precision due to the increased data volume and complexity.

\section{Conclusion}
In this paper, we extended the analysis of \model, a Topic-oriented User-Interaction model for Expert Finding in Community Question\&Answering platforms. \model integrates content and social data by constructing a Multi-Layer Graph to represent user interactions based on topical answering patterns. This approach allows \model to leverage both topic specificity and social relationships within the community to identify and rank the most knowledgeable users for any given question.
The ablation study results show that TUEF's performance is particularly notable in larger communities with well-defined topic clusters. Additionally, by incorporating interpretable Learning-to-Rank algorithms (i.e., IlMart \cite{lucchese2022ilmart}), \model achieves complete transparency, allowing a comprehensive understanding of the decision-making processes without significantly decreasing performance.
Our extensive experiments conducted on multiple Stack Exchange communities demonstrate that \model outperforms state-of-the-art EF models in both the Expert Ranking and Expert Subsample Ranking categories. It excels in precision-oriented metrics such as P@1, NDCG@3, MRR, and R@5, with improvements over the state-of-the-art by a minimum of 42.42\% in P@1, 32.73\% in NDCG@3, 21.76\% in R@5, and 29.81\% in MRR.
The study also highlighted TUEF's ability to handle larger datasets, as evidenced by its performance on datasets corresponding to 1, 2, 3, and 4 months of StackOverflow data collection. Specifically, \model consistently ranks relevant experts in top positions, demonstrating its reliability and robustness.

\begin{acks}
This work was partially supported by: the H2020 SoBig- Data++ project (\#871042); the CAMEO PRIN project (\#2022ZLL7MW) funded by the MUR; the HEU EFRA project (\#101093026) funded by the EC under the NextGeneration EU programme. A. Passarella’s and R. Perego’s work was partly funded under the PNRR - M4C2 - Investimento 1.3, PE00000013 - “FAIR” project. However, the views and opinions expressed are those of the authors only and do not necessarily reflect those of the EU or European Commission-EU. Neither the EU nor the granting authority can be held responsible for them.
\end{acks}

\bibliographystyle{ACM-Reference-Format}
\bibliography{bib}
\appendix

\end{document}